\documentclass[showpacs,preprintnumbers,amsmath,amssymb,twocolumn,superscriptaddress,aps,pre,letterpaper]{revtex4}

\usepackage{ifpdf}
\ifpdf	
	\newcommand{\ignoreThis}[1]{}
	\usepackage{color}	
	\definecolor{Gray}{rgb}{0.6,0,0}
\else		
	\newcommand{\ignoreThis}[1]{#1}
	\usepackage[usenames,dvips]{color}
\fi 
\usepackage[normalem]{ulem}

\usepackage{epsfig,graphics,graphicx,amssymb}
\usepackage{multirow}
\usepackage{framed}
\usepackage{xr}
\usepackage{url}

\begin{document}

\title{Staged Models for Interdisciplinary Research}
\author{Luis F. Lafuerza}
\affiliation{Theoretical Physics Division, School of Physics and Astronomy, The University of Manchester, Manchester M13 9PL, UK}
\author{Louise Dyson}
\affiliation{Theoretical Physics Division, School of Physics and Astronomy, The University of Manchester, Manchester M13 9PL, UK}
\affiliation{Current address: Mathematics Institute, University of Warwick, Coventry CV4 7AL, UK}
\author{Bruce Edmonds}
\affiliation{Centre for Policy Modelling, Manchester Metropolitan University, Manchester, M15 6BH, UK}
\author{Alan J. McKane}
\affiliation{Theoretical Physics Division, School of Physics and Astronomy, The University of Manchester, Manchester M13 9PL, UK}



\begin{abstract}
Modellers of complex biological or social systems are often faced with an invidious choice: to use simple models with few mechanisms that can be fully analysed, or to construct complicated models that include all the features which are thought relevant. The former ensures rigour, the latter relevance. We discuss a method that combines these two approaches, beginning with a complex model and then modelling the complicated model with simpler models. The resulting ``chain'' of models ensures some rigour and relevance. We illustrate this process on a complex model of voting intentions, constructing a reduced model which agrees well with the predictions of the full model. Experiments with variations of the simpler model yield additional insights which are hidden by the complexity of the full model. This approach facilitated collaboration between social scientists and physicists -- the complex model was specified based on the social science literature, and the simpler model constrained to agree (in core aspects) with the complicated model.
\end{abstract}

\maketitle

\medskip

{\large\textbf{Introduction}}

\medskip

To a surprising degree, the physical world can be understood through simple models (where by `simple models' we mean those that can be fully analysed). However, it is inevitable that some phenomena will not be adequately represented in this way, as seems to be the case for many biological and social systems~\cite{Edmonds2013}. In such cases, the scientist is faced with an invidious choice: to use a simple model that can be rigorously understood but does not  adequately capture the phenomena of interest; or to use a complex model that includes all the details considered necessary, but may be impossible to analyse. When trying to understand very complex phenomena, researchers from different disciplines have tended to prioritise different goals in modelling, theoretical physicists emphasizing analytical tractability and social scientists concentrating more on relevance.

\begin{figure}
 \centering
 \includegraphics[width=0.3\textwidth]{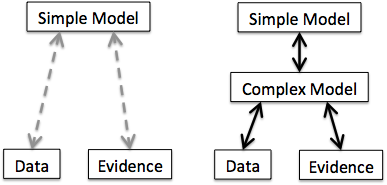}
 \caption{From a single to a multi-stage abstraction process.}\label{Fig1voter}
 \end{figure}

In this paper we suggest and demonstrate a method which attempts to combine some of the best features of both approaches. This method stages the modelling process by constructing a ``chain" of models, instead of jumping to a relatively simple model immediately (Fig. \ref{Fig1voter}). It starts with a complex but incompletely understood model and then reduces it to a simpler model that approximates some behaviours of the original. By using two, closely related, models rather than one, we hope to (a) ground the relevance of the specification of the simpler model; (b) identify key behaviours that are amenable to relatively simple representation; (c) understand the conditions under which this simplification may hold; and (d) better understand the more complex model. The disadvantages of the approach mostly relate to the increased work involved in construction and comparison.

However, such an approach is dependent on being able to simplify a complex model that one does not fully understand. In this paper, we argue that dealing with a formal, rather than a natural phenomenon, does not invalidate the normal scientific method and we show that this can be an effective approach to model simplification. In other words, to treat the complex model as if it was some natural phenomenon and to proceed to model it in the usual ways. Using this methodology, one can gain understanding of the complex model, and hence indirectly of the original natural phenomena. Furthermore, this allows additional validation of the hypothesised mechanisms, since each model can be used to check the other: hypotheses about the complex model's processes and behaviour can be explored using the simpler model, and the robustness of the simpler model probed by doing experiments (on the complex model) as to the safety of the simplifying assumptions. 

What do we mean by a ``normal scientific method"? There has been a significant amount of philosophical discussion about this, from those that think that an identifiable normative standard should be imposed~\cite{Hempel65}, to those that think that \textit{any} constraint upon method is counter-productive~\cite{Feyerabend93}. Here we mean something much more mundane. Put simply, we mean some combination of the following strategies:
\begin{enumerate}
\item observation of the target phenomena to understand its mechanisms.
\item extraction of data from the target phenomena by measurement.
\item constructing models of the target phenomena.
\item assessing models by comparing their outputs with the data.
\end{enumerate}

Each of these strategies can be equally applied to natural and formal phenomena. For example, if modelling the movement of ants one would take into account knowledge gained from observing them --- e.g. that they are social animals and might follow each other. Similarly, if modelling a simulation one would naturally inspect the code and use one's understanding of its mechanisms in a simpler model. Extracting data via measurement is much easier from a simulation than from natural phenomena since this can be done automatically using additional program code. Similarly comparing models using their output data is straightforward. 

We do not presume to specify the correct, or most effective sequence, of the above strategies. Nor do we argue here for any particular rules about how the models in strategy (3) are formulated -- whether from scratch or whether by adapting existing models. However, we do contrast this kind of approach with a purely deductive one, where one might attempt to reproduce some target phenomena using formal deduction from the formal structure of the original model.

Despite simulation reduction being a relatively under-studied issue~\cite{Chwif&al2006}, there has been some work on this within the simulation community. Here the emphasis has been mostly on complexity introduced as a by-product of simulation design and construction. For example, Innis  and Rexstad~\cite{Inn&Rex83} list 17 categories of simplification techniques, but most of these are either (sensible) engineering steps to prevent the introduction of \textit{unnecessary} complexity or seek to exploit features characteristic of particular types of systems where simplifications are possible. However, they do include sensitivity analysis to see if some variables can be omitted and ``Repro-meta modelling" --- making a model of a model (as we do here). 

Brook and Tobias~\cite{Brooks&Tob00} distinguish three kinds of model reduction: coding tricks; simplifications that preserve the output of interest exactly; and simplifications that preserve the output approximately. The first is of interest to anyone who is building a simulation---part of the range of techniques that are used to retain control over a developing complex simulation~\cite{Galan&al09}. The second seeks an exact reduction. However, this tends to destroy the meaning of the content of the models they reduce (e.g.~\cite{Thomas&al08} approximates the input-output functions implicit in a model with a neural network and~\cite{Danos&al10} simplifies by collecting model entities into abstract entities that enable a more efficient representation). It is the third category of approximating the output, that is of interest here.

There are fundamental limits to what automatic model reduction can achieve. Automatically checking whether one part of the code is functionally the same as another is, in general, undecidable (almost all general questions concerned with comparing the outcome of programs are undecidable, see any textbook on computability, e.g.~\cite{Cutland80} or read \cite{Edmonds&Bryson04} for an examination of this specific question). Thus checking whether a simulation matches its specification is also undecidable. These sorts of results (which are simple corollaries of Turing's undecidability theorem) mean that there will always be limitations to automatic techniques. This does not invalidate such methods but does imply that approaches that look for approximate and pragmatic simplifications will always be necessary. Machine learning techniques can automatically seek for representations of complex data and so could be applied to simulation outputs (final or intermediate) to infer models given their specific assumptions, but this does not result in simpler models from the point of view of a human trying to understand the dynamics \cite{Freitas14}. The models may have a more uniform structure and less complex assumptions but the results are often so complicated as to be completely opaque \cite{Burrell15}.

Most of these techniques are not aimed at distributed phenomena but at simpler targets. The work of Ibrahim~\cite{IbrahimScott04, Ibrahim05} is an exception and addresses rule-based agent-based simulation. This proposes a framework for model reduction that (a) limits the reduction to answering particular `questions' about the outcomes, (b) allows for approximate as well as exact methods, and (c) allows for a set of reduction strategies to be included. However this approach only partially works on more complex simulations. 

We cannot prove that this `normal scientific method' is a more effective way of model reduction than a deductive method. We can however, describe a case where this method was effective, that therefore supports the above approach of using a complex and a reduced simpler model in concert.

In this paper we will take as our example a complex model of voting, that considers the way various different social pressures lead to voter turnout. We first outline a complex model of voter turnout, which has been discussed elsewhere in more detail \cite{modelref,modelref2}. Next we apply the ideas discussed above to this model and obtain a reduced model, the predictions of which are then compared to the original model. Finally we conclude with a general discussion of the potential of this approach and of further work on the reduced voter model. The supporting information (SI) contains more details about the models and comparisons.

\medskip

{\large\textbf{Modelling voter turnout}}

\medskip

In this section we will outline a complex model of voter turnout which has been constructed by a group of social scientists, in collaboration with one of the authors of this paper, to encapsulate the processes that are suggested by the literature on voter turnout~\cite{modelref,modelref2}. This is sufficiently complicated that the reduction process used to create a simpler model can be appreciated. Most modelling research on voter turnout, carried out by social scientists, is based on statistical analyses; there is no tradition in constructing models of voter turnout based on the interactions of many agents. There are, nevertheless, a number of studies available that model voting behaviour as a social influence process \cite{Sznajd1,Sznajd2,Borghesi1,Borghesi2,BouchaudField,Fowler,Voterdata}. These tend to consider quite simple models that intend to capture, in a stylised manner, some aspect of the voting process or to reproduce some observed regularity.

\begin{figure}
\centering
\includegraphics[scale=0.20]{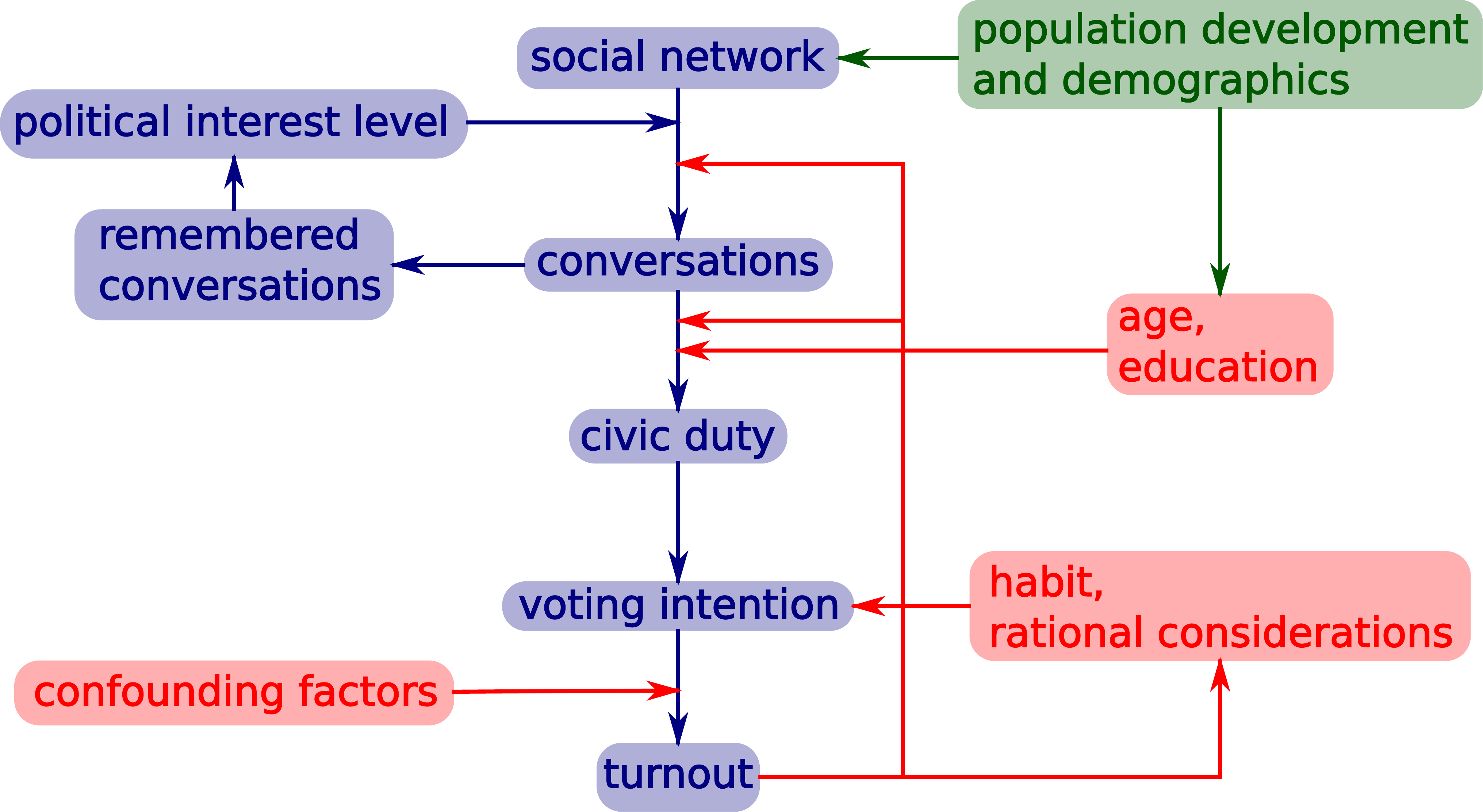}
\caption{\textbf{Diagrammatic representation of the full model processes.} The main pathway is shown in blue, with additional factors in red, and development of the agent population in green.}\label{F:OM}
\end{figure}

In contrast, we start from a complex model of voting. This model was specified by social scientists to reflect the current micro-level evidence concerning how and why people vote. Below, we give an overview of its main mechanisms. The full model can be thought of as a core structure whereby a changing population of agents generate a social network, spread influence over this network, and finally decide whether or not to vote in a general election (see Fig.~\ref{F:OM}). In addition to these basic structures, there are other sub-processes and feedbacks within these mechanisms. 

\medskip

{\textbf{Population development and demographics:}} 

\medskip

A population of agents occupy sites on a square lattice, corresponding to households, places of work or activities, schools or are simply empty. Agents have numerous characteristics (age, education-level, ethnicity \emph{etc.}), and some of these may change during the simulation. Agents are born, age and die, immigrate and emigrate. Agents that are not born in the simulation (\emph{i.e.} those initialised at the start of the simulation, or who are immigrants into the simulation), are created using demographic and socio-political data taken from the 1992 wave of the British Household Panel Study (BHPS)~\cite{BHPS}.

\medskip

{\textbf{Network formation and change:}}

\medskip

Agents create links, giving rise to several parallel social networks, corresponding to different types of relationships: partners, family, neighbours, via children at school, work colleagues and via mutual membership of activities. Partnerships and other friendships are formed primarily between agents that are `similar' in terms of age, ethnicity, class and political identification. A `friend of a friend' process occurs within each kind of network. Links can be also dropped. Agents can move house (for example when partnering) and can find and change jobs and activities.

\medskip

{\textbf{Influence spread:}}

\medskip

Agents initiate (political) conversations over these social networks (with different probabilities for different networks) with a frequency that depends on their level of interest in politics. This political interest level is determined by the number of instances of such conversations remembered. The conversations are forgotten over time with some probability. The recipient of each conversation is chosen at random from the agent's links on the corresponding network. If the agent has civic duty then the recipient may gain civic duty and may also be persuaded to change their political identification. The number of conversations children have while living with their parents strongly influences their future political interest.

\medskip

{\textbf{Decision whether to vote:}}

\medskip

Agents intend to vote, for the party they identify with, if they have civic duty, or are in the habit of voting. They may also have an intention to vote based on whether their actions in previous elections have led to a desired outcome. Agents who intend to vote may be prevented from voting by confounding factors such as illness or newborn children. Agents will acquire voting habit if they vote in three consecutive elections and will lose it if they fail to vote in two consecutive elections.

We do not have space here for a complete description of this simulation (see the supporting information for more details). This is the point---this simulation is too complicated to completely understand, being formed of a complex mix of social processes that affect each other. Rather in this paper, we aim to describe how we sought to understand this by modelling it with simpler models, in a manner very similar to that if the target of analysis had been some natural phenomena. If our target for modelling had been some natural phenomena we would only had been able to give a similarly brief sketch of relevant aspects.

\medskip

{\large\textbf{Model reduction and comparison}}

\medskip

To understand the complex model we constructed a series of reduced models, and then compared them to the original and each other. The procedure consisted of: creating reduced models (strategy 3, in the above) by removing or approximating aspects of the model that we expect to be less important; comparing the output with that of the original model (strategies 2 and 4); formulating hypotheses as to the origins of discrepancies between the models, based on careful observation and analysis of the mechanisms involved (strategy 1); iterating this procedure with new models. 

After several iterations of of the above processes we found a reduced model that gave a sufficiently good fit in terms of outputs as compared with the original. The aspects of the original model which were removed or approximated are:
\begin{enumerate}
\item[] \textit{Social network}: As a first approximation, we initially removed the social network, so that each individual may talk to any other agent in the simulation.
\item[] \textit{Political parties}: Since we focus on turnout, we are not directly interested in which party was voted for. In the full model this does have an indirect effect on turnout, since agents may choose not to vote due to ``rational considerations'', dependent on their judgement about the efficacy of their past history of voting on the desired outcome of the election. However, we did not expect this to play a large role in whether agents vote and hence removed parties and such ``rational considerations'' by agents.
\item[] \textit{Children}: Since only adults vote, we eliminated the explicit raising of children. Instead we approximated this part of the model by creating all new agents as 18-year-old adults, with characteristics assigned to them based on approximations of the full model. 
\item[] \textit{Confounding factors}: In the full model agents that intend to vote may not vote due to a number of confounding factors (such as recent unemployment, illness, young children, etc.). We replaced these with a general probability of not voting, which is a function only of age. 
\item[] \textit{Population size}: We used a fixed population size, so that every emigration event is matched by an immigration event, and every death by a birth.
\end{enumerate}

\medskip

{\textbf{Comparison between the reduced model and the full model (M$_1$ with M$_2$)}}

\medskip

Although we compared the reduced and full model using several measurements, we will focus here on the dynamics of turnout, i.e. the proportion of the population who vote, since this was the principal focus when building the full model. In particular, we examine the range of voter turnouts for different values of the parameter `influence rate', which scales the average number of times per year that agents initiate conversations.

\begin{figure}
\centering
\includegraphics[width=0.35\textwidth]{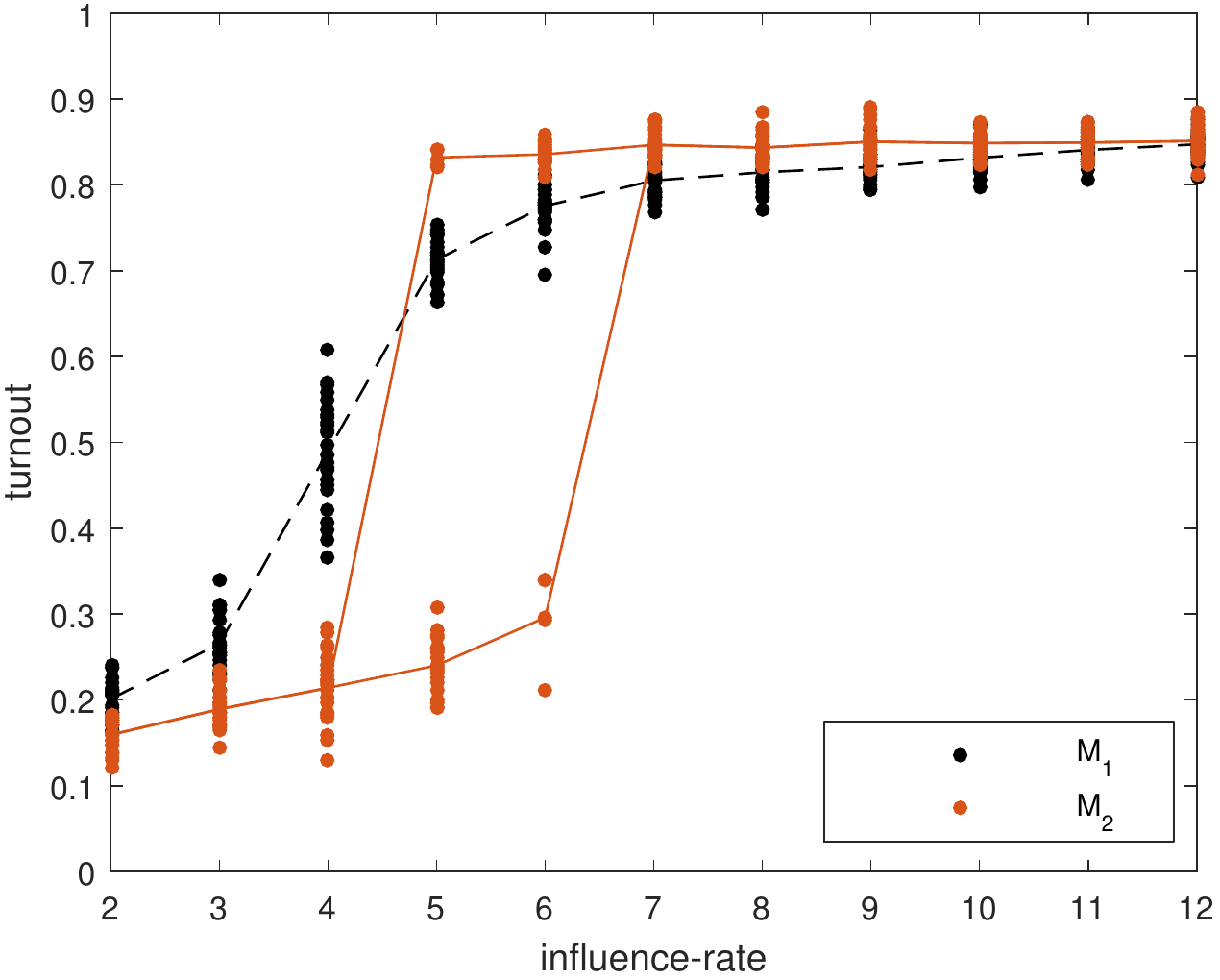}
\caption{\textbf{Comparison between the full model (black) and the reduced model (red).} Ten different values of the influence rate parameter (from 2 to 11) are shown. For each one, the steady state value of the turnout obtained is shown for 25 realisations (dots), together with the mean values (lines).}\label{F:M1_vs_M2}
\end{figure}

We first compare the reduced model (denoted M$_2$) described above with the full model (M$_1$). Since we have removed many mechanisms we do not expect to obtain full agreement between the two models but since we retain the most vital parts of the full model we do expect to see some qualitative agreement. This comparison is shown in Fig.~\ref{F:M1_vs_M2}.

Both models have two main `modes': a high-turnout mode, corresponding to a high average number of conversations; and a low-turnout mode, corresponding to a lower number of conversations. The existence of different modes was only discovered through considering the reduced model, as it is over 1000 times faster to run. The two modes are due to the following: civic duty (acquired by talking to other agents that have civic duty) is highly correlated with voting, so that an agent with civic duty is very likely to vote; in parallel to this when agents are spoken to this increases their interest in politics, which then increases their likelihood of speaking to other agents. Thus increasing the influence rate parameter increases the overall amount of conversations in the model, both directly, through the effect of the parameter value, and indirectly, since agents that receive conversations are also more likely to initiate conversations. This feedback loop (Fig.~\ref{F:OM}) amplifies the effect of the parameter and hence `locks-in' a high level of turnout. Conversely, low civic duty levels and likelihood of conversations have the opposite effect.

We also find some quantitative agreement: in the high-turnout mode both models predict that roughly 80\% of the electorate vote; while in the low-turnout mode voting is at around 20\% (although the full model gives a somewhat higher level in the low-turnout regime). The parameter values in the reduced model are determined directly from the original model, and are not the result of `fitting' the output. In the reduced model we find that for some intermediate values of the influence rate, the same initial conditions and parameter values can lead to either a high-turnout or a low-turnout mode in different simulations (bistability). In contrast, in the full model there is no region of bistability, instead intermediate influence rates lead to intermediate levels of voting. 

The existence of high- and low-turnout regimes may be of practical interest. It suggests that the effect of efforts to increase voting might strongly depend on the parameter regime of the system. If we are close to the transition from low-turnout to high-turnout, then a small increase in the number of conversations people have about politics could be amplified to give a large effect on turnout, disproportionate to the initial effort of increasing the number of conversations. Conversely, if we are far from the region of transition, either deep in the low-turnout regime or in the high-turnout regime, then efforts to increase voting by increasing how much people speak about politics may have little effect.

From this point on, we add additional mechanisms to the reduced model and compare the outcomes to determine the effects and importance of these mechanisms. Each new mechanism will be described in the following sections and in the supporting information, and will be given an acronym to distinguish the different versions of the reduced model. For example, the reduced model with the `clumped network' (described below) will be referred to as M$_2$+CN.

\medskip

{\textbf{Adding a synthetic ``clumped'' social network removes the bistability (M$_2$ vs. M$_2$+CN)}}

\medskip

In the full model there are three networks generated by the model, each one used to carry out political conversations with a different probability. The most used network describes partnerships between couples, who live together on a single lattice site, perhaps with children or other adults. The second most used network depicts family relationships, which are typically between individuals that live on the same lattice site. Usually, each family is completely internally connected. Finally, there are long-range links, used less frequently, which give all other sorts of friendships (\emph{e.g.} neighbours, work colleagues, school-friends \emph{etc.}). Each of these networks are dynamic, evolving over time as agents age and change. The development of the networks includes mechanisms that take into account the homophily of the agents, so that agents are more likely to make friends with those who are most similar in terms of class, ethnicity, education level and political views.  In contrast, in the reduced model agents talk with a randomly selected other agent in the model. This is equivalent to having a fully connected social network

\begin{figure}
\centering
\includegraphics[width=0.35\columnwidth]{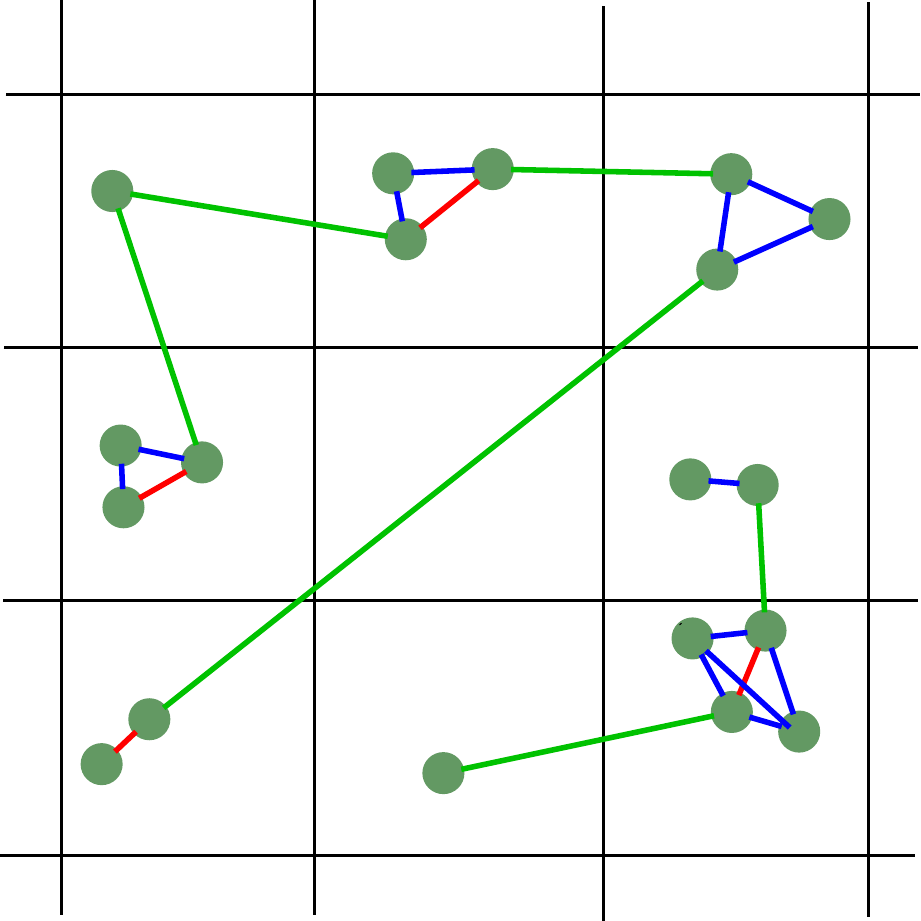}
\includegraphics[width=0.35\columnwidth]{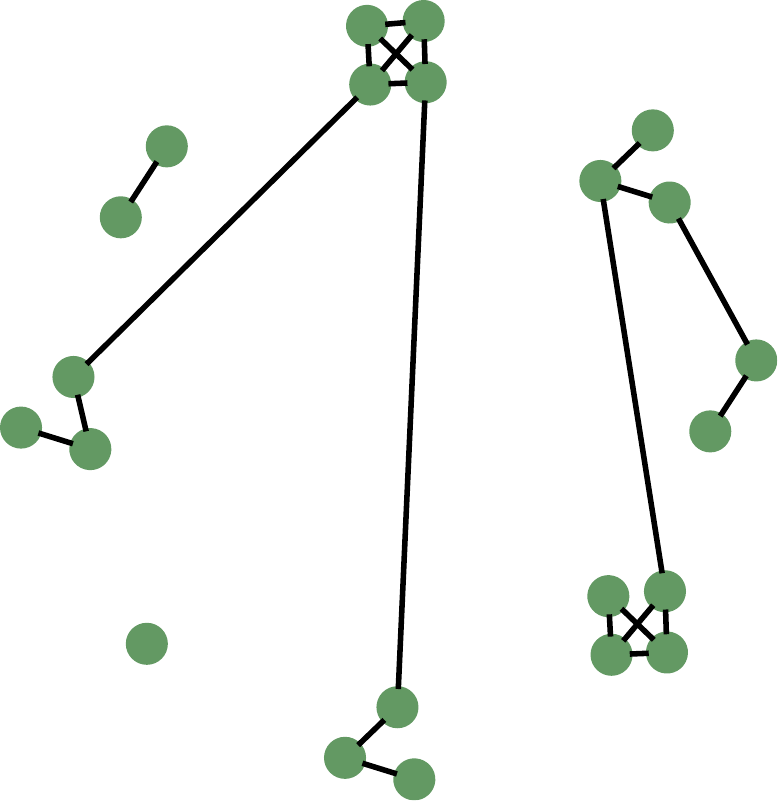}
\caption{\textbf{Schematic comparison between the full model network (left) and the synthetic network (network CN, right).} Agents are displayed as green circles. Lines connecting agents represent social links. In the full model red lines represent partners, blue lines represent families and green lines represent other kinds of relationships.}\label{F:network_comparison}
\end{figure}

We hypothesise that the important features of the network generated by the full model, with regards to turnout dynamics, relate to the general structure of the network, and not to specific characteristics of individual agents. To test this hypothesis we make a synthetic `clumped' network (network CN) by creating totally connected groups of agents (representing households) and then rewiring some of these internal connections to create a few long-range links between groups (see Fig.~\ref{F:network_comparison}). Thus there are two parameters for creating network CN: the average degree of an individual, and the probability of rewiring each link. A higher average degree creates a network that is more similar to that used by the reduced model, where agents may talk with any other agent. A higher rewiring probability makes the network more similar to a randomly connected network. It reduces the clustering~\cite{WattsBook} of the network, and represents the situation where households are less important compared to friendships outside the family group. The size of the initially fully connected groups was taken to be a uniform (discrete) random variable between 1 and 8, and each link was rewired with a probability of 0.12. In this way the degree distribution, the clustering coefficient and the proportion of local to long-range links were similar to those obtained in the full model (a precise fitting of these quantities is problematic since we are comparing a multi-level network to a simple one). This network CN is closely related to the so-called ``caveman'' graph~\cite{WattsBook} -- it is a network that consists of ``clumps'' of well connected nodes with some long-range links between them.

\begin{figure}
\centering
\includegraphics[width=0.35\textwidth]{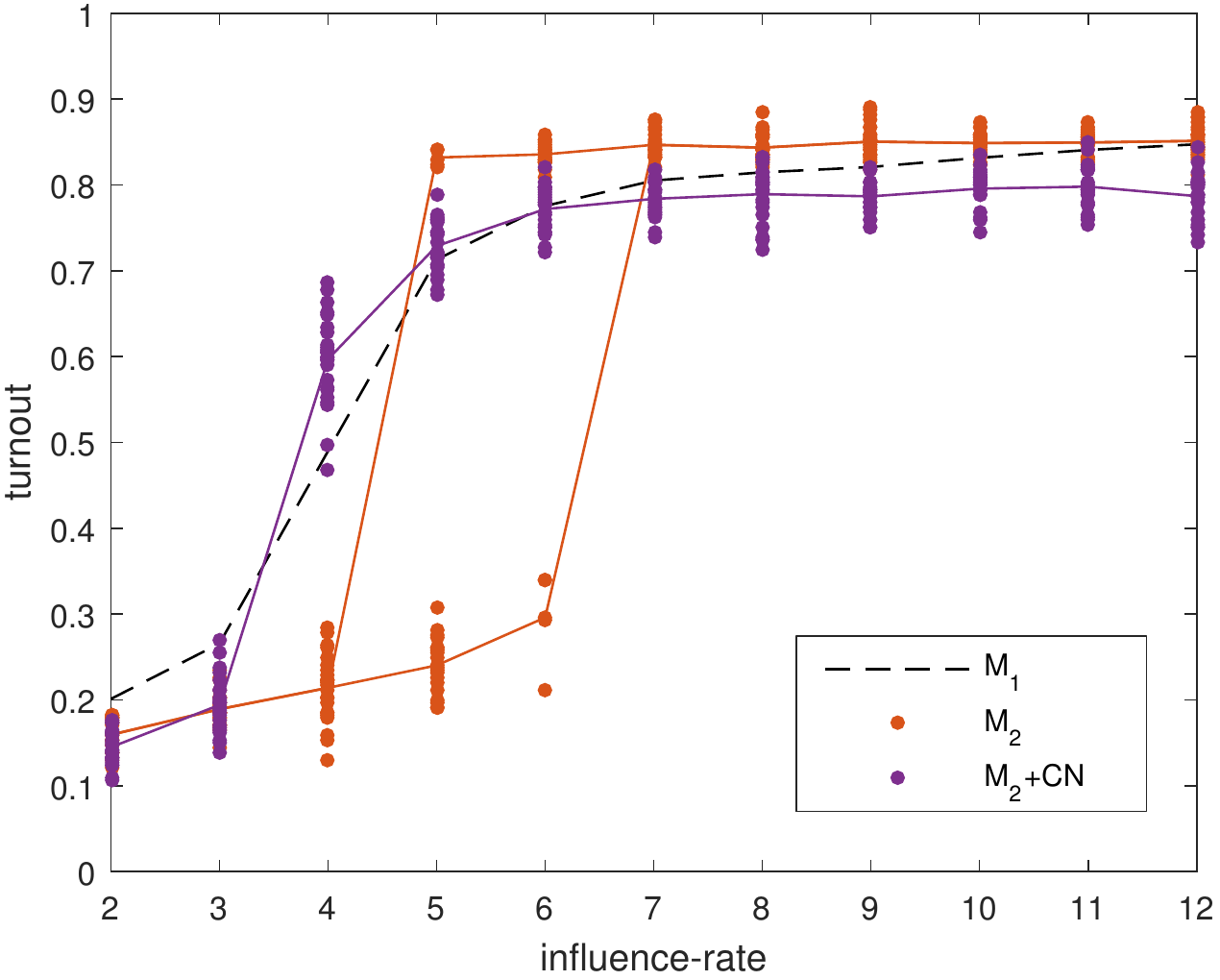}
\caption{\textbf{Comparison between the full model M$_1$, (dashed black), the reduced model, M$_2$ (red) and version M$_2$+CN (purple).} The synthetic network (network CN, see the main text) decreases turnout in the high-turnout regime and leads to a transition between low- and high-turnout at a lower influence rate.}\label{F:M1_vs_M2_vs_M2CN}
\end{figure}

The reduced model but with this network (M$_2$+CN) gives a turnout similar to that of the full model. In particular, the turnout is slightly lower than without the network, and the transition from low- to high-turnout occurs at a similar value of the influence rate to the full model (Fig.~\ref{F:M1_vs_M2_vs_M2CN}). 
Importantly, the bistability obtained with the fully connected network disappears. This network effect is due to the sparseness and the `locality' of the network, and is further explored in the SI.


\medskip

{\textbf{Making the network dynamic leads to higher voting (M$_2$+CN vs M$_2$+CN+D)}}

\medskip

\begin{figure}
\centering
\includegraphics[width=0.35\textwidth]{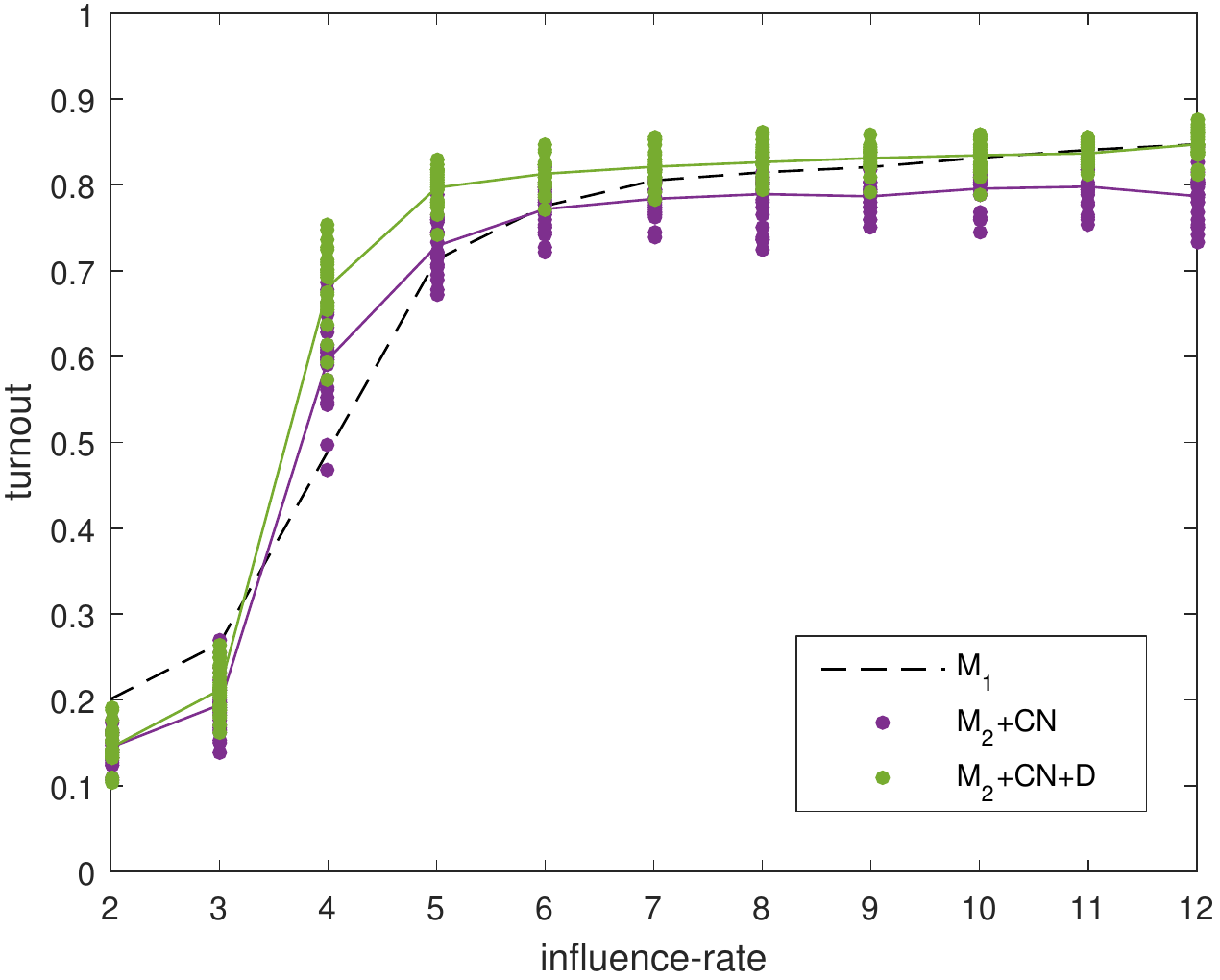}
\caption{\textbf{Comparison between the full model, version M$_2$+CN (purple) and version M$_2$+CN+D (green).} Rewiring the network (with a probability of 0.15 per long-range link per year) increases turnout.}\label{F:M1_vs_M2CN_vs_M2CND}
\end{figure}

Using a synthetic social network (network CN) replicates the more gradual transition between low- and high-turnout regimes found in the full model, removing the bistability seen in the reduced model. However the turnout seen in the high-turnout regime is still lower than that found in the full model. One potential reason for this discrepancy is that the reduced model has a static network. To test this, we include dynamic network rewiring into the model from the previous section (M$_2$+CN+D) in the following way. Every year, each long-range link is broken with probability 0.15 and one end of this link is reconnected to another agent selected at random. When we compare this to the model without dynamic rewiring (M$_2$+CN), we find that the turnout is indeed increased in the high-turnout regime (Fig.~\ref{F:M1_vs_M2CN_vs_M2CND}). In the low-turnout regime using a dynamic network does not significantly affect the turnout, since the probability that a low-interest agent is rewired to be connected to a high-interest agent is very low anyway. The increase in turnout due to dynamical rewiring was larger when the underlying network was sparse (network CN has an average degree of 3.5). The effect of dynamic rewiring is related to that found in \cite{Centola07}, where rewiring facilitated the achievement of  consensus in an Axelrod model~\cite{Axelrod}. Despite the improvement in agreement with the full model, the increased turnout in the high-turnout regime sometimes over-predicts turnout (Fig.~\ref{F:M1_vs_M2CN_vs_M2CND}), indicating that additional mechanisms should be considered. 

\medskip

{\textbf{Immigration by household leads to lower turnout than immigration by individual (M$_2$+CN vs M$_2$+CN+HI)}}

\medskip

The last aspect we will consider is the implementation of immigration. 
Immigration in the model is rather high, with around half of the adult agents having been introduced as immigrants (note that immigrants correspond to all the agents not born in the population, not just foreigners). For this reason, we can expect that the particular form of immigration that we use will affect the results. In the reduced model each agent may emigrate with a given probability, and is replaced with a new immigrant agent. In contrast, agents in the full model immigrate and emigrate as households, so that all new immigrant agents begin the simulation living with other immigrant agents. This has the biggest effect when immigrants are drawn from a population that is significantly different from the native population. 

\begin{figure}
\centering
\includegraphics[width=0.35\textwidth]{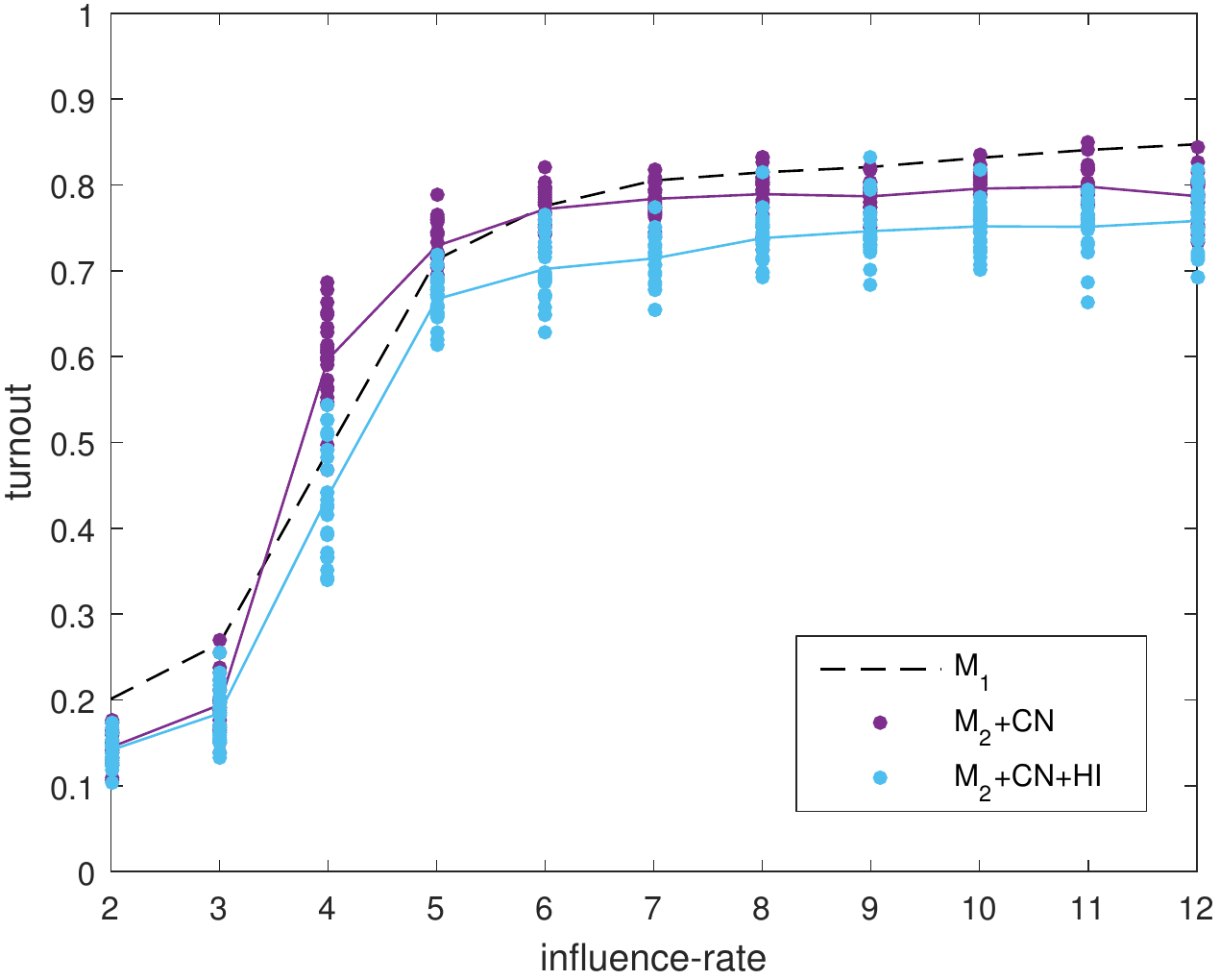}
\caption{\textbf{Comparison between version M$_2$+CN (purple) and version M$_2$+CN+HI (light blue).} When immigration occurs by households turnout is reduced.}\label{F:M1_vs_M2CN_vs_M2CNHI}
\end{figure}

We compare the reduced model with the synthetic network (M$_2$+CN) where immigration occurs individually to the same model but where immigration occurs by household (M$_2$+CN+HI). This comparison is shown in Fig.~\ref{F:M1_vs_M2CN_vs_M2CNHI}.  We see that this does indeed lower the level of turnout. This can be explained as follows. In the high-turnout regime, where agents native to the model are very likely to have civic duty and high interest in politics, new immigrant agents are likely to have lower interest in politics and are less likely to have civic duty in comparison to the native population. If these new agents immigrate as a household, then the majority of their social links will be within that group, so that it is unlikely that an interested agent will initiate conversations with them. Thus agents forming a household of immigrants with low interest in politics are unlikely to increase their interest through conversations with native agents. Conversely, if agents immigrate individually and integrate into established groups of native agents, then these agents are likely to have higher interest in politics, and will initiate conversations with the immigrating agents, who will increase their own interest in politics. We therefore expect that including immigration by households in the high turnout regime will lead to moderately lower interest in politics and hence lower turnout overall due to lower turnout in the immigrant population. Thus using immigration by individual increases the effect that the higher-talking immigrants can have on the current native population. This mechanism is illustrated in the SI.

Thus if one wishes to increase turnout in the general population, our model indicates that it is always better if immigrant agents are individually integrated into the general population. This result depends on the one-way nature of the influence included in the model, where agents can make others more interested in politics (and thus more likely to initiate conversations about politics), or teach others to have civic duty, but cannot make other agents less interested or less likely to have civic duty. While this is an assumption of the model, it is grounded on some evidence~\cite{Rolfe2012}.

\medskip

{\textbf{Including both extensions results in a better fit with the full model (M$_2$+CN+D+HI vs M$_1$)}}

\medskip

\begin{figure}
\centering
\includegraphics[width=0.35\textwidth]{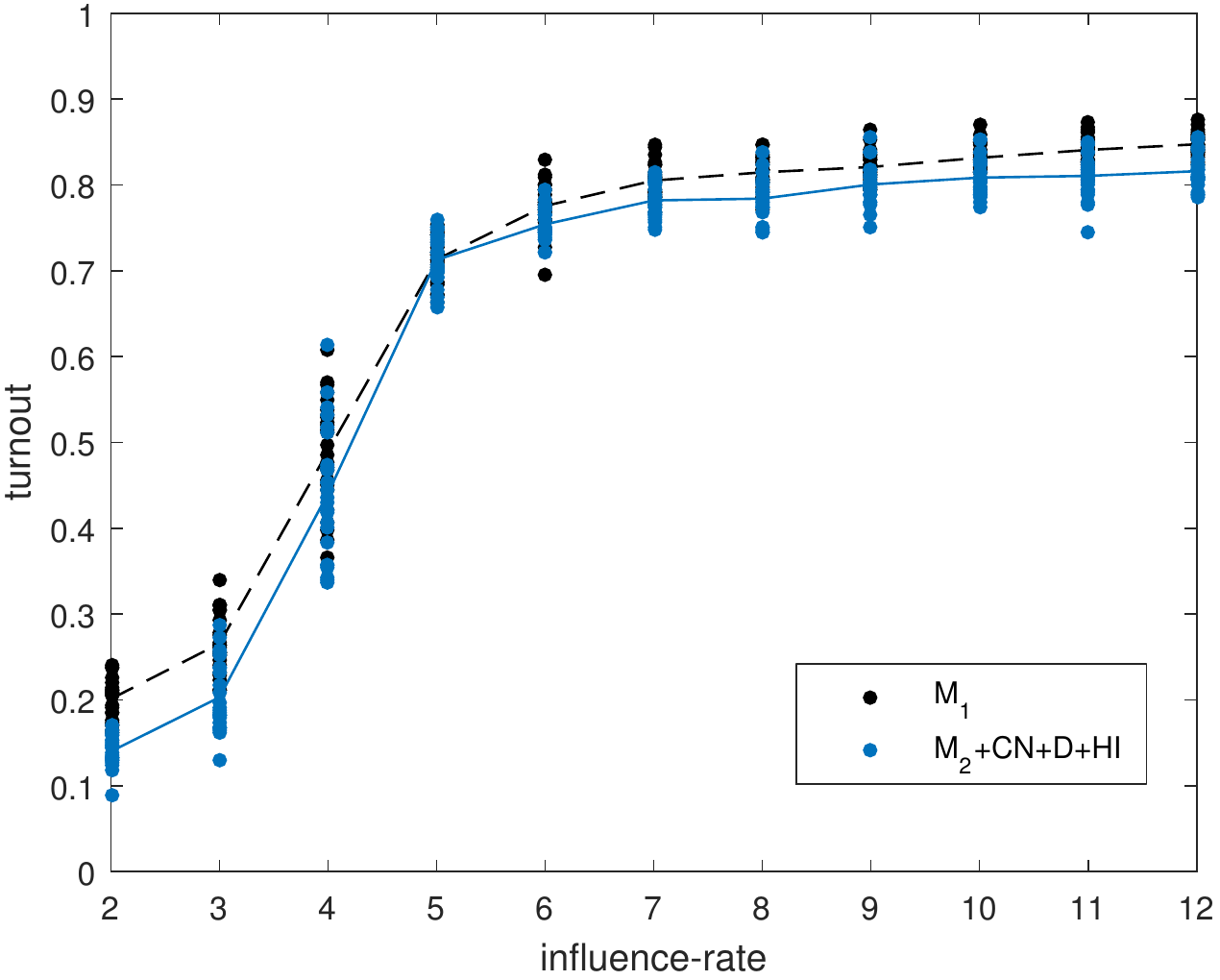}
\caption{\textbf{Comparison between original model (black) and version M$_2$+CN+D+HI (blue).} Ten different values of the influence rate parameter (from 2 to 12) are shown. For each one, the steady state value of the turnout obtained is shown for 25 realisations (dots), together with the mean values (lines).
}\label{F:M1_vs_M2CNDHI}
\end{figure}

In the preceding sections we discussed the impact of two mechanisms: allowing dynamic network rewiring during the simulation; and implementing  immigration by households. Here we include both of these mechanisms simultaneously (M$_2$+CN+D+HI) and demonstrate that with both of these mechanisms the model compares well with the full model at all values of the influence rate parameter (Fig.~\ref{F:M1_vs_M2CNDHI}).  Thus we have substantially improved the fit of the reduced model for the particular target of turnout dynamics and for the parameter ranges considered. It might be that different reduced models will be suitable for different output targets and parameter ranges.

Although version M$_2$+CN+D+HI includes more mechanisms than the fully reduced model, M$_2$, it is still significantly simpler than the full model. Version M$_2$+CN+D+HI runs approximately 1000 times faster than the full model (see the supporting information). Exploration and analysis of the reduced model is substantially easier and faster than with the full model, although we have maintained a relatively close correspondence between the two.

\medskip

{\large\textbf{Discussion}}

\medskip

In this paper we have described a method for simplifying and understanding a complex model using a series of simpler models, and have used this method to better understand an intricate agent-based model of voting. We have thus demonstrated the effects of the different mechanisms included in the model, and have shown that significant simplifications can be made without compromising the target results over a particular range of parameter values. We believe that this approach can be applied to the analysis of other complex phenomena that cannot be adequately represented using a single simple model. The approach detailed here provides a structure to facilitate interdisciplinary collaborations between data-driven modelling, requiring a high level of detail, and an analytical approach, requiring simpler models that are more amenable to systematic analysis. In this way, insights obtained in the simpler models can be seen to be relevant to the more complicated models, and the systems they describe.

We stress that our model-reduction approach involves no substantial data fitting. Most of the parameter values of the reduced model are directly given by those of the full model. When there is not a direct correspondence of parameters, such as for parameters of the reduced network (which is a simple network as opposed to the multi-level character of the full model's network), the parameter values are chosen so that the microscopic characteristics (such as degree distribution or proportion of long-range links) mimic those of the full model. 

The kind of process described above has resulted in simpler models that are different than one might invent in a one-step modelling process---the composition of these models was guided by what was in the complex model and what turned out to be significant for the narrow question of turnout dynamics. The simpler models have resulted in insights into the workings of the complex model---insights that would have been difficult to obtain through direct simulations due to the slowness of execution of the complex model and its very large parameter space. Thus the importance of the detail of the social network has been revealed for the transition between high-turnout and low-turnout regimes, and the potential different impact of different modes of immigration highlighted. The process of developing all these models and comparing different variations is relatively time consuming, but one ends up with a chain of related models that combine some of the advantages of simplicity with the assurance of relevance. 
We have recently developed a voter model which is even simpler than the one presented here and which is amenable to a mathematical analysis. This was constructed as a further reduced version of $M_2$ described in this paper. This then is the next link in the chain, and will be discussed in detail elsewhere.

\bigskip

We thank Ed Fieldhouse, Laurence Lessard-Phillips, and others in the SCID project for useful discussions. This work was supported by the EPSRC under grant number EP/H02171X.

\appendix

\vspace{1cm}

{\large\textbf{SUPPORTING INFORMATION}}

\bigskip

{\large\textbf{Further details on the model analysis}}

\medskip

{\textbf{Effects of a static network}}

\medskip

\begin{figure}[!h]
\centering
\includegraphics[width=0.35\textwidth]{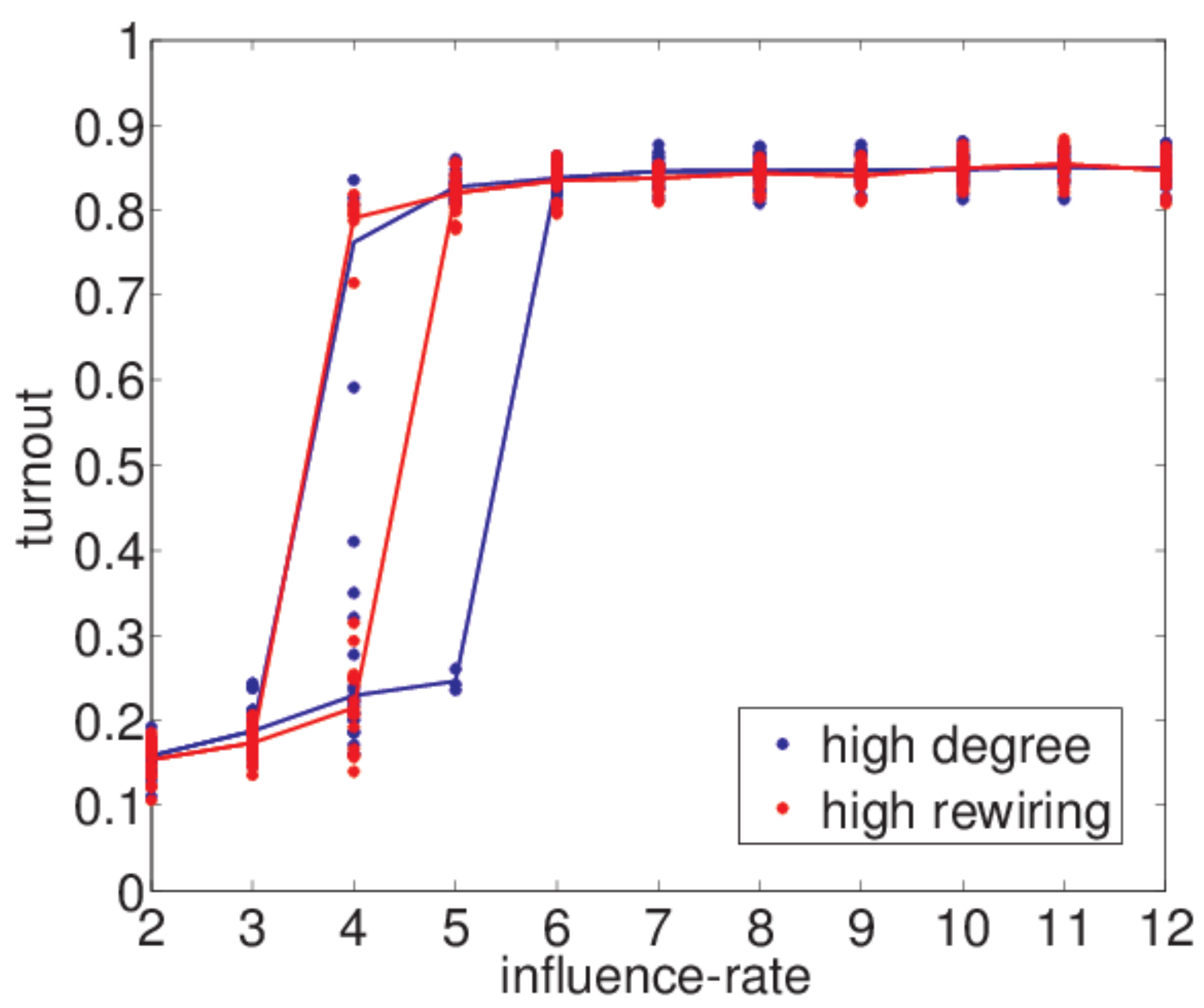}
\caption{\textbf{Using a network with either high degree (blue) or high rewiring probability (red) regains the bistability observed in the reduced model.} Both simulations use model $M_2$. The high degree network has degree = 65 and rewiring probability = 0.15. The high rewiring network has degree = 15 and rewiring probability = 0.45.}\label{Model1_beta_bistability}      
\end{figure}

In the main text we show that, while a fully connected network leads to a sharp transition with a region of bistability, a (fixed) network made of small, strongly connected communities with a few links between them leads to a smoother transition in which the bistability region disappears (see Fig.~5 of the main text). More precisely, the network considered was made of fully (internally) connected groups whose size was taken to be a uniform (discrete) random variable between 1 and 8; then, each link was rewired with a probability of 0.12. More generally, we can consider a family of networks of this type parametrised by the average size of the groups, $s$, and the rewiring probability, $r$. Here $s$ controls the average degree (or, equivalently, the network density or connectivity), and $r$ the degree of `clumpiness' (related to the modularity and the clustering coefficient). Using this family of networks, we can explore the effects of the connectivity and the `clumpiness' on the dynamics. We find that when $s$ or $r$ are increased, the dynamical results are more similar to those obtained for the fully connected network, and the bistability region is recovered, as illustrated in Fig.~\ref{Model1_beta_bistability}. In this way, it is clear that the connectivity and the `clumpiness' of the network are crucial characteristics in determining the behaviour of the model, particularly in the region of intermediate social influence. This prediction was confirmed in the full model, $M_1$, where it was found that artificially increasing the number of links could give rise to bistable behaviour.

\medskip

{\textbf{Effects of the immigration implementation}}

\medskip

\begin{figure}[!h]
\centering
\includegraphics[width=0.35\textwidth]{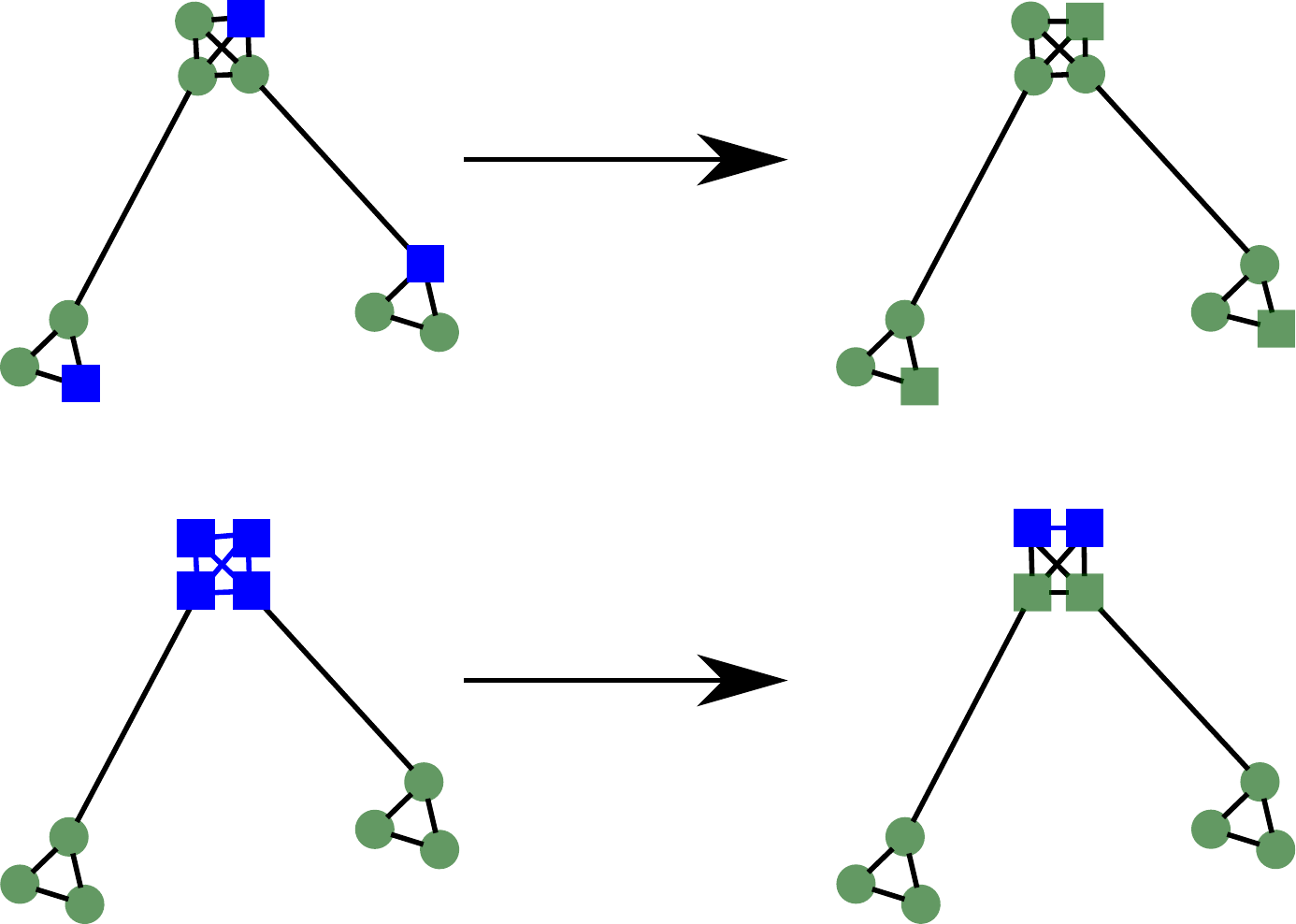}
\hfill
\includegraphics[width=0.35\textwidth]{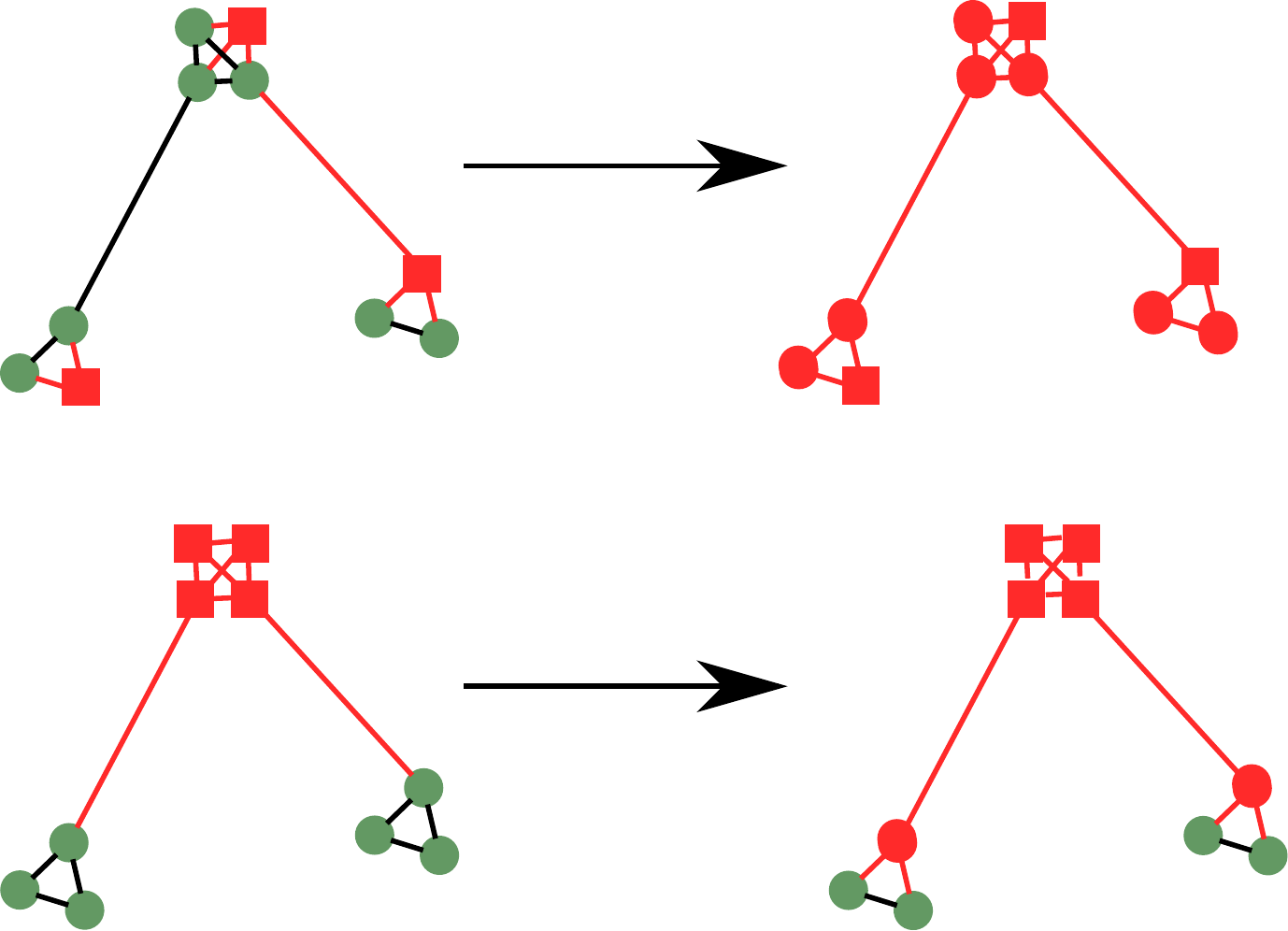}
\caption{\textbf{A schematic representation of the effect of immigration by individual vs. immigration by household:} (left) when immigrants have a lower political interest and a lower chance of having civic duty than the general population and (right) when immigrants have a higher political interest and a higher chance of having civic duty than the general population. In each diagram, native agents are represented by circles and immigrants by squares. Political interest is shown in red (high interest), green (normal interest) and blue (low interest). In both cases immigration by individual leads to a higher interest in the population.}\label{F:immigrant_dynamics}
\end{figure}

In the main text we show that when immigrants enter the simulation via whole households the overall turnout decreases (compared to the case in which immigrants enter individually). The mechanism giving rise to this result is illustrated in Fig.~\ref{F:immigrant_dynamics}. The origin of this effect lies in the asymmetry of the social influence process, whereby highly interested individuals can increase the interest of less interested individuals but less interested individuals cannot decrease the interest of highly interested individuals. Under these circumstances, a higher level of overall interest is achieved when highly interested individuals are more connected to less interested individuals, while a lower level of interest is achieved if highly interested individuals are more connected with other highly interested individuals. This is so because the influence of highly interested individuals is `wasted' if they talk to one another, while it has a larger effect if it is concentrated on less interested individuals. 
When the immigrants are different to the incumbent population, it will boost turnout if they tend to be connected to the incumbent population; if immigrants have higher interest they will be able to pass this interest to the rest of the population; if immigrants have lower interest they will have the chance to increase it via contacts with the rest of the population.  

\medskip

{\textbf{Runtime comparison}}

\medskip

\begin{figure}[!h]
\centering
\includegraphics[width=0.35\textwidth]{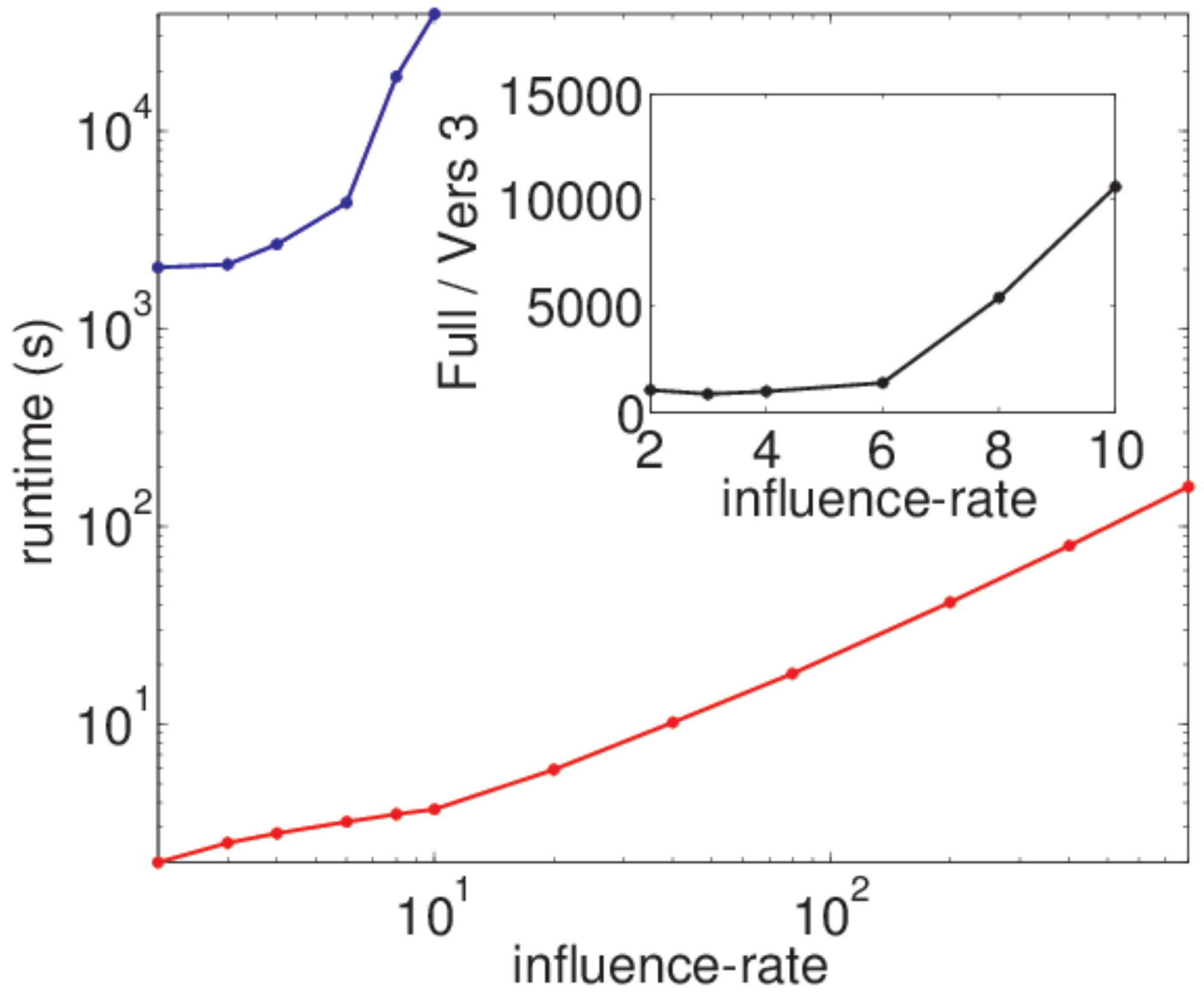}
\caption{\textbf{Run time comparison between the original model $M_1$ (blue) and the reduced version $M_2+CN+D+HI$ (red).} The time taken to run the simulation for different values of the influence-rate parameter for the original model, $M_1$, (blue) and version $M_2+CN+D+HI$  of the reduced model (red). Note that for values of influence-rate above 10 simulations of the original model did not successfully complete. The inset shows the quotient of the two simulation times. Simulations were performed on a standard desktop computer.}\label{F:Runtime_comparison}
\end{figure}

To assess the difference in computational demands of the full and the reduced model, we compared the (real) time needed to run the models on a standard desktop computer. We used the parameter values employed in the main text, as indicated in Table \ref{parametervalues}, varying the parameter influence rate, which controls the overall number of conversations in the population. We compared the full model, $M_1$, with version $M_2+CN+D+HI$  of the reduced model, which is the more complex of the reduced models. The reduced model was implemented in C programming language (while the full model is written in NetLogo). The results are shown in Fig.~\ref{F:Runtime_comparison}.

\medskip

{\large\textbf{Reduced model description}}

\medskip

Each agent (with index $i = 1, \ldots, N$) has the following list of characteristics, some of which may change over time:
\begin{itemize}
\item[] \textbf{binary variables:} civic duty ($CD(i)$), turnout (in last election, $v(i)$), habit ($h(i)$), post-18 education ($e(i)$);
\item[] \textbf{integer variables:} (political) interest level ($l(i)$), minimum interest level ($m(i)$), age (in years, $a(i)$), number of remembered (political) conversations ($c(i)$).
\end{itemize}

The main parameters of the model are:\newline
Influence rate $K$, scales the number of (political) conversations per year.\newline
Probabilities of initiating a conversation $\mathbf{p}_c(l,v)$.\newline
Probabilities of gaining and loosing civic duty.\newline
Thresholds on the number of conversations needed to increase the interest level $Th_{\alpha}$.\newline
Probability of forgetting a conversation $p_f(l)$.\newline
Death probability $p_d(a)$.\newline
Emigration probability $p_e$.\newline
Probability of not voting due to confounding factors $p_c(a)$.

\medskip

{\textbf{Initialisation procedures}}

\medskip

Agents are initialised using data from the British Household Panel Study (BHPS) \cite{BHPS}. The same procedure initialises immigrants into the model, using the subset of the BHPS corresponding to survey responses from immigrants. This procedure sets the civic duty, turnout, habit, post-18 education, interest level and minimum interest level, with some of these characteristics being inferred using proxies for the required information. Agents initially have no remembered conversations, and an age drawn from a uniform distribution between 18 and 70 (to initialise the model) and between 18 and 48 (for later immigrants into the model). Agents born during the simulation are initialised at age 18 (we do not explicitly model their growth until age 18), and education with probability 0.3. Their interest level and minimum interest level is equal to their education, and they are assumed have no civic duty, habit or remembered conversations and not to have voted in the last election.

\medskip

{\textbf{Main loop}}

\medskip

The following processes happen in a loop until the required timepoint is reached. All rates are given in Table \ref{parametervalues}.
\textbf{Each year:} 
	\begin{itemize}
		\item[] \textbf{Each month:}
		\begin{itemize}
			\item[] \textbf{Carrying out conversations:} For each agent, this section is run $\lfloor $K$ / 12 \rfloor$ times plus one time extra with probability $ $K$ / 12 - \lfloor $K$ / 12 \rfloor$. 
			\begin{itemize}
				\item[] The agent has the chance to initiate three conversations, with probabilities $\mathbf{p}_c(l(i),v(i))$  each with a random other agent.
				\item[] Agents (with $l(i)>0$) receiving a conversation (from an agent with civic duty), acquire civic duty with probability $p_{acd}(e(i),v(i))$.
			\end{itemize} 
			\item[] \textbf{Updating interest levels:} 
			\begin{itemize}
				\item[] If $l(i)=0$ and $c(i)>Th_0$ then set $l(i) = 1$ and $m(i) = 1$.
				\item[] Else, if $c(i)>Th_h$ then set $l(i) = m(i) + 2$.
				\item[] Else, if $c(i)>Th_l$ then set $l(i) = m(i) + 1$.
			\end{itemize}
			\item[] \textbf{Updating civic duty:} Agents lose civic duty with probability, $p_{lcd}(a(i),e(i))$, dependent on their age and education.
		\end{itemize}
		\item[] \textbf{Forgetting conversations:} Agents forget conversations that happened more than one year ago, with probability, $p_f(l(i))$, per conversation, dependent on the agent's interest level.
		\item[] \textbf{Birth/death:} Each agent dies with a probability, $p_d(a(i))$, dependent on their age, and is replaced by a new agent by the `birth' process (described in the Initialisation procedures).
		\item[] \textbf{Immigration/emigration:} Each agent emigrates with a probability $p_e = 0.015$ and is replaced by a new agent by the `immigration' process (described in the Initialisation procedures).
		\item[] \textbf{Ageing:} Agents age by one year
	\end{itemize}
\textbf{Every 5 years there is an election:}
	\begin{itemize}
		\item[] Agents with civic duty or habit vote unless `confounded' (due to illness or other factors) with probability $p_c(a(i))$, dependent on their age.
		\item[] Agents gain habit if they vote in 3 consecutive elections.
		\item[] Agents lose habit if they do not vote in 2 consecutive elections.
	\end{itemize}
Here $\lfloor x\rfloor$ denotes the integer part of $x$, that is, the largest integer less than or equal to $x$.
\begin{table*}[!ht]
\caption{
{\bf Parameter values of model $M_2$.}}\label{parametervalues}
\begin{tabular}{l|c|r}
\hline
\textbf{Parameter}         & \textbf{Value}                            & \textbf{Meaning}                                                                                                \cr
\textbf{name} &&\cr \hline
N                               &  480                                      & population size                                                                                        \cr \hline
$p_d(a)$                        & a function of age derived from mortality tables	& death rate                                                                                             \cr \hline
$p_e$                           & 0.015                                     & emigration rate                                                                                        \cr \hline
$K$                             & $K \in [2,12]$                            & influence rate                                                                                         \cr \hline
\multirow{6}{*}{$p_c(l,v)$}     & $\mathbf{p}_c(2,0) = [0.0100,0.0500,0.1500]$    & \cr
                                & $\mathbf{p}_c(2,1) = [0.0600,0.1000,0.1800]$     &                                                                                                        \cr
                                & $\mathbf{p}_c(3,0) = [0.0600,0.1925,0.3795]$& probability of initiating a conversation \cr
                                & $\mathbf{p}_c(3,1) = [0.1540,0.2800,0.3900]$  &                                                                                                       ($\mathbf{p}_c(l,v) = [0,0,0]$ if $l\leq 1$) \cr
                                & $\mathbf{p}_c(4,0) = [0.2000,0.4750,0.5134]$  &                                                                                                         \cr
                                & $\mathbf{p}_c(4,1) = [0.3232,0.5680,0.5370]$&                                                                                                        \cr \hline
$p_acd(e,v)$                    & 1-(1-0.25(1+e))(1+v))(1-0.125(1+e))(1+v)) & probability of acquiring civic duty                                                                    \cr \hline
$Th_0$                          & 5                                         & threshold for increasing interest level to 1                                                           \cr \hline
$Th_l$                          & 2                                         & lower threshold for increasing interest                                                                \cr \hline
$Th_h$                          & 5                                         & higher threshold for increasing interest                                                               \cr \hline
\multirow{2}{*}{$p_{lcd}(a,e)$} & $0.01/(12(1+e))$ if $a\geq 25$           & \multirow{2}{*}{probability (per month) of losing civic duty}                                                      \cr \cline{2-2}
                                & $0$ if $a < 25$                           &                                                                                                        \cr \hline
\multirow{2}{*}{$p_c(a)$}       &  $0.077$ if $a\leq 75$                    & \multirow{2}{*}{probability of not voting due to being confounded}                        \cr \cline{2-2}
                                & $0.077+(1-0.077)0.9^{(a-75)(a-74)/2}$  else   &                                                                                                        \cr \hline
\multirow{2}{*}{$p_f(l)$}       & 0.2 if $l = 0$                            & \multirow{2}{*}{probability (per year) of forgetting a conversation}                                              \cr \cline{2-2}
                                & 0.5 if $l \geq 1$                         &                                                                                                        \cr \hline
\end{tabular}
\end{table*}

\medskip

{\large\textbf{Description of the Full Model}}

\medskip

Here we give more detail about the full model ($M_1$). This description will follow the ``ODD'' protocol for this \cite{Grimm06}.  The full code, a complete description of the details of the model and a sensitivity analysis can be found at \cite{modelref}. 

\medskip

{\textbf{Overview}}

\medskip

\textit{Purpose of Model}

\medskip

This is intended as a consistent, detailed and dynamic description, in the form of an agent-based simulation, of the available evidence concerning the question of why people bother to vote.  This integrates a variety of kinds and qualities of evidence, from source data and statistics to more qualitative evidence in the form of interviews.  The model is being developed following a KIDS rather than a KISS methodology, that is, it aims to be more guided by the available evidence rather than simplicity \cite{Edmondskisskidd}.

\medskip

\textit{Entities, state variables, scales}

\medskip

\begin{figure}
 \centering
\includegraphics[width=0.6\textwidth]{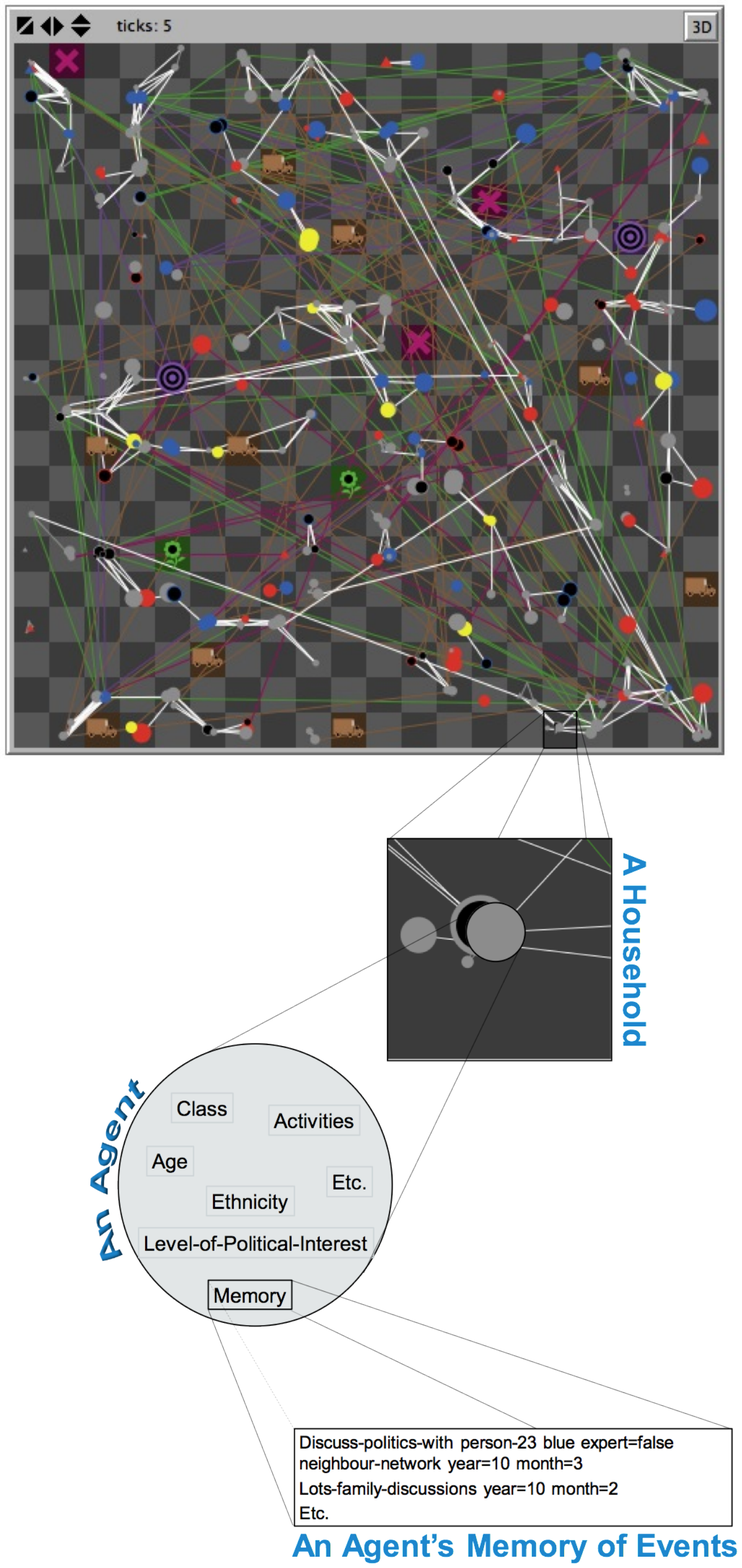}
 \caption{\textbf{An illustration of the model elements}. Each patch represents a household or other location (place of work, school, or kind of activity). The circles or triangles on patches are the agents of a household. The links between them social networks of different types. Agents have a number of attributes, including: their ethnicity (shape), their political leaning (colour), and their age (size). Other agent attributes include class, level of political interest, and which activities they belong to. They also have a (partial) memory of past events, including who they voted for and the political discussions they have had.}
\label{SI:Fig1}
\end{figure}

The model is based around a 2D grid of locations, each of which may be a: household, place of work, school, activity (two kinds) or empty.  Households consist of a number of agents which each represent a single person.  Agents are born, age, partner, have children and die as the simulation progresses.  Agents have a large number of characteristics, but these include: a memory of past events, a party affiliation (or none), a set of family relationships (children, partner, and/or parents) and social connections with other agents.  It is over the network of social relationships that influence occurs in the form of events that represent communication about political or civic matters.  The agents are influenced over time via these communications.  When an election occurs, these influences, along with other factors, affect whether an agent votes and, if so, for which party.  The collected votes are endogenously summed to give the election result, which might affect whether agents consider voting again. These elements are illustrated in Fig.~\ref{SI:Fig1}.

Places of work, schools and activities are place-holders.  They do not change or move (unlike the households).  Their only characteristic is their membership (who works there, which children go to school there, which are members of an activity).  A household is simply a container for the agents who form that household.

Agents are the primary elements in the simulation and have many characteristics, including:

\begin{itemize}
\item Age: the number of simulation years since they were born
\item Ethnicity: their ethnicity (majority, invisible minority, or visible majority)
\item Partner: the agent who is their partner (if any)
\item Children: their children (if any) 
\item Older-relations: adults in their household when born (if any)
\item Parents: their parents (if any)
\item Employed?: whether they are employed
\item Last-lost-job: when they lost their last job (if unemployed)
\item Voted?: whether they voted in the last election
\item Voted-for: who they voted for in the last election (if any)
\item Year-last-child: when they had their last child (if any)
\item Immigrant-gen: a number indicating if they are direct immigrants (0), the next generation (1), etc.
\item Class: Which of the 5 classes the agent belongs
\item Moved-out?: whether the agent has left their birth home (ever)
\item Ill?: whether the agent is currently ill
\item Post-18-edu?: whether the agent has acquired a post-18 education
\item Civic-duty?: whether the agent has a sense of civic duty (with respect to voting)
\item Interest-level: the level of interest an agent has in politics, one of: not noticing politics, noticing politics, taking a view on issues, politically interested, and politically involved
\item Party-habit?: coming from voting for a party in consecutive elections
\item Gen-habit?: whether the agent has the habit of voting
\item The number of consecutive elections in which the agent has voted
\item The number of consecutive elections in which the agent has voted for the same party
\item The number of consecutive elections in which the agent has not voted
\item Memberships: those schools, places of work or activities that an agent belongs to
\item Social links: those other agents with whom they might (if both were inclined) have a political discussion
\item Memory of events: such as recent political conversations they have had, whether they felt satisfied with voting, or whether there was lots of political discussion at home
Other characteristics exist for the purpose of simplifying programming, debugging and collecting statistics and these are not listed.  Note that currently agents are not sexed so partnerships can be formed between any two agents.
\end{itemize}

\medskip

\textit{Process overview, scheduling}

\medskip

The simulation is initialised at the start.  Then the simulation proceeds in discrete time steps, one step usually representing each month in a year.  Each time step the following stages are carried out.

\begin{enumerate}
\item External immigration and households moving into area from outside of the UK sampled from immigrants in BHPS sample, unless grid is full
\item Internal immigration and households moving into area from inside of the UK sampled from all BHPS sample, unless grid is full (remixed in terms of given majority/minority mix)
\item Emigration and households moving out of the area
\item Birth and death probabilistically using statistics
\item Forgetting and things being lost from the endorsements of agents at different rates, e.g. remembrance of conversations
\item Network-changes and social links to other agents and activities made and broken
\item Partnerships are formed, move to live together if possible
\item Partnerships dissolved, one partner moves out
\item Household might move within the simulation area
\item Have conversations and hold conversations over the social network, influencing others in the process, the frequency of this is adjusted using the influence-rate parameter
\item Updating agents' attributes in terms of: noticing politics, interest level, and civic duty
\item Once a year update: the party preference, party habit and generalised habit
\item Drift-process and shift of voters into and from each political party by a drift process: voters for ruling party (not very interested in politics) drift away to grey, some grey drift to a party
\item (During the short run-up to an election): politically involved agents probibilistically contact random adult agents and have a political conversation with them which may increase their probability to vote.
\item During an election tick: 
\begin{itemize}
\item Determine an agent's intention to vote based upon: their satisfaction with their past experience of voting, whether they have a sense of civic duty, whether they have acquired the habit of voting, whether they feel an identification with a particular party, their level of interest in politics
\item Factor in some of the effects of confounding factors (such as having recently moved, or having a child under 1) and record some statistics about consecutive voting etc.
\item If it is to occur (determined by the simulation settings) the effect of party mobilisation efforts are computed
\item For all those going to vote, then may drag others to vote with them
\item The voting process itself happens including recording information of who voted and why (for the model analysis)
\item The election result is determined via endogenous voting by agents (although it could be exogenously fixed if this is set)
\end{itemize}
\item Updating various plots and statistics for output about what is happening in the simulation
\end{enumerate}

For each of these stages agents are fired in a random order (newly random each time and process).  In most of these processes the update for each agent has no immediate effect on any other agent, so these agent processes are effectively in parallel.  Similarly most of these stages could be done in any order with very little impact on the outcome, the exception being the sub-stages of voting (item 14 above).

\medskip

{\textbf{Design}}

\medskip

\textit{Basic principles}

\medskip

The starting point for the model design was a collection of 54 ``causal stories'' about behaviour that might be relevant to whether people bother to go and vote.  Each such story traces a single causal thread through the complexity of social and cognitive processes whilst letting the context of these be implicit and whilst ignoring their possible myriad interactions.  This ``menu'' of behaviours drove the architecture of the model as it was designed to allow most of these stories to be explicitly represented.  When the simulation is run, the local conditions of each agent separately define the context of that agent whilst also allowing the complex mixing of many different social and cognitive processes.  

To fill in some of the cognitive and contextual ``glue'', evidence from many different sources has been included to motivate the assumptions and mechanisms of the model.  Thus it is difficult to identify discrete ``sub-models'' in this.  However, a post-hoc analysis of the structure that emerged suggests the following could be considered as sub-models:

\begin{enumerate}
\item The main social unit is the household, a collection of individuals living within the same house.  People who partner may form a new household, or people moving from outside the area may also do so.  Many social processes occur within the household and others occur preferentially between members of the same household.  Households occupy a place on the available square 2D grid.
\item Basic demographic processes specify how people enter the model (though moving into it from the UK or abroad), are born to partners, age, leave home, partner, separate, and die. These are based on some available statistics as to the probability of these events, depending upon whether the immediate situation of the agent makes these plausible (e.g. no one gives birth without a partner, nobody can separate until partnered etc.).  This demographic model includes a 5-category social class model, with statistics from class mobility determining when and where people move class.
\item To this basic demographic is added a number of activities.  These are currently schools, places of work, activity type1 and activity type2 (these may be thought of as to correspond to things like: places of worship, sports clubs etc.).  Agents between the age of 4 and 18 attend school; those 18-65 can go to a place of work, and join (or leave) activities.  The activities are recorded as `place-holders' on the grid of households and take up a location but they have no characteristics except their current membership.  All children are members of the nearest school; if in work adults are members of a random place of work; with a certain probability adults join an activity and, if so join the one whose other members are (on average) most similar to themselves.
\item A dynamic social network develops between agents.  Each link represents a relationship that would allow for a conversation about politics and voting should the participants be up for this.  The links are typed and the types are: partner, household, neighbourhood, work, school, activity1 and activity2.  There are several different ways that a new link can form: all people in the same household are linked with a household link, there is a chance that people in neighbouring households might link with a neighbouring link, people who go to the same school or parents of children who go to the same school might link with a school link; people who are members of the same activity might form a link.  Further for each of these link types there is a chance of making a link with someone linked to someone an individual is linked to (``friend of a friend'').  Links can be dropped under certain circumstances and with certain situations (e.g. if one moves, most of the neighbourhood links are lost).
\item Agents can have different levels of political interest (from lowest to highest): not noticing politics; noticing politics; taking a political view on issues; interested in politics, involved in politics. They also have other associated attributes, such as (possibly) a: party political leaning (the party they would vote for if they did), a sense of civic duty to vote, a generalised habit to vote, a party identification, and a memory of whether past voting/not brought about their desired outcome.
\item A process of social influence occurs over this social network in the form of discrete (as opposed to continuous) political discussions.  A political discussion occurs if: (a) there is a link between the two (b) the talker is at least interested in politics (has at least a view on politics) and (c) the receiver at least notices political discussions (there is a lower level of awareness that occurs in the home and elsewhere to get people up to the level of noticing political discussions).
\item These political discussions have several possible effects (when taken in aggregate): they may increase the level of political interest of the listener, they may help impact a sense of civic duty and they may help convince the listener to adopt a political leaning.  There are some slow processes whereby these may be forgotten over time.
\item When an election occurs, each individual goes through a process which determines whether they vote or not: (1) if they have a sense of civic duty, or general voting habit (2) rational calculations such as whether the balance of past voting experiences was positive and whether they have a strong party identification (3) (if this occurs) political parties may mobilise some who have leanings towards them but were not intending to vote (4) positive intentions to vote may be confounded by factors such as: have a very young baby, having just moved, having just been made unemployed or being to ill to vote, (5) finally those going to vote may ``drag'' others to come with them and vote, especially partners or family.
\item Voting statistics are then recorded, with agents remembering where and when they voted, with the election result being decided by the majority vote within the model (although an option is that it could be imposed from outside).
\end{enumerate}

The above are not the full details but a summary of their main features. Generally micro-causation in the model happens down the order above (from first to later), but there are some weaker and slower feedbacks that occur back up, for example the outcome of an election effects agents' perceptions of the experience of voting (whether voting resulted in the party they wanted); the characteristics of agents (including party leaning) may affect which activity they join, their friends and who they choose as a partner; and (most importantly) political discussions affect the level of interest of agents.

\medskip

\textit{Emergence}

\medskip

Clearly in such a complicated model it is not possible to make an easy and clean distinction between results that emerge and those that are programmed into the model.  Indeed, the model was designed with a view to integrate available evidence rather than produce or demonstrate emergent effects (or to be predictable).  However it is not the case that all outcomes from the model are straightforwardly forced by the settings and programmed micro-processes, including the following.

\begin{itemize}
\item Although the underlying demographic model is fairly predictable in its unfolding, which partnerships are formed affect which new households with children are created (that do not result from people moving into the district from outside), so the developing social network affects the demographics a little.
\item The patterning of households within the 2D space has certain self-organising features. Households have a tendency to move to districts where surrounding households will have some similar agents to themselves, resulting in some weak clustering. The positioning of schools also has an effect as children will go to the nearest school, and links may be formed between parents of children at the same school.
\item Agents will tend to choose to participate in (voluntary) activities whose other members are (on average) most similar to themselves, so that these activities tend to act to promote clustering of similar individuals, regardless of location.
\item Depending on the network structure, clusters of agents will tend to reinforce patterns of interest/lack of interest in politics.  This may reinforce or act against tendencies that might already exist within households of different kinds within the simulations (which will for the reasons above tend to cluster together in terms of location and activity membership etc.).
\end{itemize}

The initialisation of the model (see below) has a complicated but predictable effect on the model, in that the kinds of household the model is seeded with will affect the tendencies that follow.  Thus in the data set that these are selected (at random) from those from ``invisible minorities'' (Irish etc.) tend to be more politically involved and have a higher sense of civic duty than the native majority population, so if the model is selected to have more of this kind one will find a higher level of turnout.

The impact of many of the parameters is straightforward, for example: increasing the probability of holding a conversation increases the general level of political interest and hence the turnout; increasing the forgetting rate (the ``forget-mult'' parameter) means that people do not recall so many positive political messages and hence the level of interest in politics falls quicker.  The immediate effect of mobilisation is fairly straightforward and the more people are mobilised the more vote -- but how this effects the longer term is less obvious in that it seems to have greatest impact upon the levels of civic duty and general habit, than (for example) in terms of a cascade effect in bringing yet others out to vote.

\medskip

\textit{Adaptation}

\medskip

Agents generally do not seek to increase or optimise any measure of success nor do they reproduce behaviours that they perceive as successful. The exceptions are: (a) when agents weigh up their past experiences of voting as one factor in the decision of whether to vote again, (b) when moving to a new location within the model, the choice might be influenced in the sense of seeking a location with neighbours similar to themselves and (c) if choosing to join a type of activity agents will choose the instance of the activity whose membership is, on average, the most similar to themselves.

\medskip

\textit{Objectives}

\medskip

Agents do not aim to meet any objective.

\medskip

\textit{Learning}

\medskip

Agents do learn, adapting their traits over time depending on their circumstances and history.

\begin{itemize} 
\item \textit{Level of interest in politics}: this is influenced by many factors, including: amount of political discussion in the parental home, whether they have had a post-18 education, and the level of experienced political discussion once an agent has left home (this needs to be higher than that within the home for the same effect).  However level of experienced political discussion only has this effect once an agent gets to the level of noticing politics (or above) which may be triggered by a certain level of discussion in the home, or a much higher level outside. 

\item \textit{Social network}: agents develop their social network in a number of ways over time: (a) they are automatically linked to other members of the same household, (b) they connect with a probability to those at the same school (or other parents with children at the same school), activity, workplace or immediate neighbours (but preferentially to those more similar to themselves) and (c) for each kind of link (neighbourhood, school, activity1, activity2, workplace) they can make a link to some of those linked to those they are linked to (so called `friend of a friend').  There is a fixed probability of dropping links at each time click, also if an agent moves they are almost certain to lose existing school, neighbourhood and household links (though there is a small probability of retaining them).

\item \textit{Political leaning}: If agents are sufficiently interested in politics, then they can be persuaded to adopt a political leaning in the following conditions: (a) in the home adopt the party of the most politically interested parent, or if both equally interested the party if they agree on this (b) outside the home change from grey (no party) to the most frequently mentioned party in political discussions it has heard, depending on the proportion of discussions for the most mentioned party and its current level of political interest (c) if the political interest level of the agent falls to below `noticing politics' then they lose their political leaning.

\item \textit{Whether they feel a sense of civic duty to vote}: political conversations that are conducted by an agent with civic duty can impart civic duty to another agent.  

\item \textit{Whether they have picked up the simple habit of voting}: people acquire a habit of voting when they have voted in 3 consecutive elections.  If they fail to vote in 2 consecutive elections they lose this habit.

\item \textit{Whether they have developed an identification with a particular political party}: if agents have voted in the previous 3 elections for the same party, they acquire an identification with that party.  If their politics ever drops to grey (no party) they lose this.

\item (During the short run-up to an election) \textit{The level of intention to vote}: at the start of the (short) campaign this will be set for each agent according to a number of factors (whether they have civic duty, have developed a voting habit, have the highest level of interest in politics, have a strong party identification and are statisfied with past voting outcomes, are a loyal supporter and are statisfied with past voting outcomes). As the short campiagn develops conversations might also have the effect of increasing this intention to vote (Depending on the level of intention in both agents).
\end{itemize} 

\medskip

\textit{Prediction}

\medskip
	
Agents do not do any prediction in this model. In particular, in this version of the model, there is no tactical voting, nor expectations about whether it is worth voting based on predicted outcome.

\medskip

\textit{Sensing}

\medskip

This is a social model, so that agents primarily sense other agents in three ways: (a) through their current links to other agents, (b) through indirect links to other agents, e.g. by being members of the same activity, having kids at the same school or being in neighbouring cells (c) through political discussions over the direct links.  Thus all sensing is local in the sense of their links, memberships or neighbourhood (except that agents are aware of the result of elections).  

\medskip

\textit{Interaction}

\medskip

Agents interact with each other by having political ``conversations'', which may influence the recipient.  Each ``conversation'' carries messages of political leaning and civic duty (depending on the characteristics of the converser).  These are not strictly conversations since each one is one way, but over time these may go both ways between agents, reinforcing existing characteristics of leaning, political interest and sense of civic duty.  If an agent moves location, it will bring its partner and children with it (as well as possibly orphaned children in the household).  Agents form sexual partnerships, selecting from those in their social network, and can only have children when within such a partnership.  Partnerships dissolve with a low random probability in which case one partner will move out leaving any children behind.

\medskip

\textit{Stochasticity}

\medskip

Many processes in the model have a stochastic element in them once the conditions for their occurrence are locally met in an agent.  This includes the processes of: moving location, emigrating, immigrating, getting a job, losing a job, making new social links or losing them, joining an activity or leaving one, having a political conversation, acquiring civic duty as a result of a conversation, dragging others to go and vote if they are going, and mobilising voters. Other process have a probability of occurring but with the probability varying on the basis of some statistics, including: birth, death, moving out of the parental home, becoming ill, and children changing class later in life from that they were born with (which also depends on having a post-18 education).

The processes that determine the probability of someone voting are deterministic but somewhat complicated (see 8 in the section on design principles and 14 under the section on scheduling).  Many circumstances, such as having a sense of civic duty or being politically involved force a probability 1 of voting (unless a confounding factor intervenes).

Processes that are entirely deterministic include: going to school or leaving it, retiring from work, the election result, changes in the habit of voting, or political identification.

A major stochastic impact on the model is in the initialisation of the households at the start of the simulation and the choice of new households that enter during the simulation due to immigration.  In these processes entire households are selected at random from re-mixed sample of households from the 1992 wave of the BHPS.  The ``re-mixing'' is done to achieve the user defined proportion of majority population as well as to ensure that out-of-UK immigration is selected from those recorded as immigrants in the BHPS sample.  Thus the mix of initial households in each run of the simulation will be somewhat different, but on the whole, the balance of household characteristics will be representative for simulations with larger populations albeit with some stochastic variation.

\medskip

\textit{Collectives}

\medskip

Some of the agent characteristics do influence how the agents make links and move. Which locations a household moves to is influenced by a bias towards moving next to households with similar characteristics; which instance of a kind of activity 1/2 are joined will be those whose existing members have (on average) the most similar characteristics as themselves; which person they make links with via an activity will be biased by a similar homophily formula.  Thus over time agents will tend to have more links with those similar to themselves.  However due to the presence of much stochasticity in the model this does not produce pronounced segregation, but rather a ``softer'' bias in terms of social links.  The characteristics that are involved are: age, ethnicity, class and political leaning.  At the moment there is a single dissimilarity measure used between two agents regardless of the context (in future versions this will be changed so that there are different measure for different circumstances, so (for example) a weaker one at work than for choosing which instance of an activity to join).

Political parties are not currently represented, except implicitly in terms of the mobilisation process. Individuals influence each other individually and not collectively in this model.

\medskip

\textit{Observation}

\medskip

Many different statistics are collected from the simulation. Broadly the more complex a simulation, the more different aspects need to be validated in order to have any confidence that the model represents what one intends it to.  Following the process of cross-validation \cite{Moss05} broad evidence and statistics are used to inform the specification micro-level agent rules but then the results coming out of the model also checked, both statistically and in broader qualitative terms. We will not describe all these here. More details can be found in the documents archived with \cite{modelref}. These include output statistics, graphs, histograms, a visualisation of the world with the social netoworks and agents shown, and there is a trace, where the events that occur to a randomly chosen agent are logged.  When this agent dies a new born agent is chosen and logged.  This is to give a feel for the sort of life trajectories agents are going through.

\medskip

{\textbf{Details}}

\medskip

\textit{Initialization}

\medskip

The grid is initialised in the following manner:
\begin{itemize}
\item The grid dimensions are set by the programmer
\item Set proportions of the grid are occupied with schools, work places, activity1 and activity2 (with a minimum of one each)
\item A given proportion of patches that are left are populated with new households.  These are selected as a complete household from a large sample of taken at random from the 1992 wave of the British Household Panel Survey (BHPS)  \cite{BHPS}, but `remixed' to a set degree of majority population (by splitting the original file into majority/non-majority households and then probabilistically choosing at random from each part according to parameter settings).  Some details about households (e.g. which child in a household belongs to which parent) have to be inferred from the data as this is not always unambiguous.  Some initial agent characteristics are set using proxies from the data, e.g. civic duty is set for agents who are recorded as being a member of certain kinds of organisation
\item Links to household members and some random neighbours are made
\item To give the households an initial network the procedure to develop other network links is done 10 times for each household.
\item Appropriate activities are joined depending on those in the BHPS data.
Thus the exact composition of the grid varies in each run but are drawn from the same sample, so in a sufficiently large initial set of households (determined by the size of the grid and how much is left empty) one gets a similar mixture each time.
Various other things are initialised including: shapes and colours for main display, election dates, and party labels.
\end{itemize}

\medskip

\textit{Input Data}

\medskip

There are two sets of data that are used in the model:
\begin{itemize}
\item A sample of the 1992 wave of the BHPS data as described above.  This file cannot be distributed due to UK Data Archive restrictions and it will be soon available on their site.  In its stead we are distributing the model with synthetic data which does not relate to any real individuals but has some of the same characteristics as the original file \cite{modelref}.
\item Various statistics concerning the underlying demographics, such as birth rate (depending on the age of parent), death probability (each age), probability of males and females leaving home.  At the moment these are statistics from only roughly the appropriate time.
\end{itemize}

\medskip

\textit{Submodels}

\medskip

It is important to understand that this is \textit{not} a simulation with free-parameters that are conditioned on some ``in-sample'' data.  It does have a lot of parameters, but these are set (or could be set) from empirical data. The model is then run ``as is'' and can be compared with available data --- to see how and where it matches this and when it does not.  Thus (unlike many models) it is not an attempt to `fit' any data, but rather is a computational description to enable the `detangling' and critique of various explanations of observed social behaviour.  

Some of the principal parameters that have real referents (that is, in principle they could be determined from empirical data), include the following:
\begin{itemize}
\item drop-friend-prob: the probability a link is dropped in a year
\item drop-activity-prob: the probability an activity membership (not work or school) is dropped each year
\item prob-partner: the probability of forming a sexual partnership if single per year
\item prob-move-near: when a household moves this is the probability it moves to the nearest empty patch rather than to a patch with similar neighbours to itself
\item immigration-rate: percentage of population that immigrates from outside the UK into the model (and hence is randomly selected from the immigrants section of the BHPS file)
\item int-immigration-rate: percentage of population that immigrates from inside the UK into the model (and hence is randomly selected from the re-mixed version of the BHPS file)
\item emigration-rate: the rate (per year) that households leave the model
\item dissim-of-empty: when judging if a neighbourhood contains similar households to self, this is how dissimilar an empty space is (thus a low value of this results in housholds seeking to move near empty spaces, a high value to avoid empty spaces)
\item election-mobilisation-rate: the percentage of its supporters who are not intending to vote that a party tries to get to vote
\item start-mobilisation: when party mobilisation starts
\item end-mobilisation: when party mobilisation stops
\end{itemize}

The following allow the turning on and off of various processes or structures and thus allows the comparison of the simulation behaviour with and without them.
\begin{itemize}
\item household-drag?: whether agents attempt to drag others to vote
\item rand-convs?: if on means that political conversations happen at random and are not constrained by the social network
\item p2p-influence?: switches whether the specific influence between discussants during the election period on their intention to vote can occur
\item no-rat-voting?: turns off the calculative (or ``rational'') aspects of the decision whether to vote
\item greys-vote?: whether those with no political inclination can vote (if they do they do so randomly)
\item mob-once-ph?: whether mobilisation conversations only occur once to each household
\item fof?: switches the friend-of-a-friend social link creation mechanism
\end{itemize}

Some of the other parameters can be used to implicitly switch processes on and off:
\begin{itemize}
\item influence rate: setting this to zero switches off all political conversation (apart from mobilisation conversations)
\item prob-contacted: setting this to zero switches off mobilisation during elections
\item major-election-period and minor-election-period: setting these to zero switches off elections
\item immigration-rate and int-immigration-rate: setting these to zero switches off any incomers to model (warning may critically affect longer-term population levels)
\item emmigration-rate: setting this to zero switches off any emigration model (warning may critically affect longer-term population levels)
\item birth-mult: setting this to zero switches off any births (warning may critically affect longer-term population levels)
\item death-mult: setting this to zero switches off any deaths (warning may critically affect longer-term population levels)
\item prob-partner: setting this to zero switches off any partnering after initialisation (warning may critically affect longer-term population levels)
\item separate-prob: setting this to zero switches off any separation of partners (warning may critically affect longer-term population levels)
\item forget-mult: setting this to zero switches off any forgetting of conversations etc. by agents (warning will cause model to slow down as agent accumulate huge lists of memories)
\item move-prob-mult: setting this to zero switches off any moving within model 
\end{itemize}

The following affect the initialisation of the simulation.
\begin{itemize}
\item density: the initial density of households in the spaces left for them after schools etc. have been allocated
\item majority-prop: the proportion of the initial population from the majority group
\item init-move-prob: how many times households are moved in the initialisation (this produces a slightly more realistic starting point for the model with weak clustering)
\end{itemize}

The following control how the simulation run occurs and what data is output.
\begin{itemize}
\item start-date: year simulation starts
\item end-date: year simulation finishes
\item ticks-per-year: how many simulation ticks are in each year and probabilities throughout the simulation are adjusted so that roughly the same will happen with different settings of this, so as to enable fast debugging runs with 1 tick per year before slower ones with 12.  However there will be subtle differences in model behaviour for different settings of this.
\item to-file?: switches whether simulation saves statistics to the file given in ``output-filename''
\item when-calc-data?: determines when the simulation saves statistics and/or network data (1=every tick, 2=every two ticks, etc.)
\item sna-out?: switches whether the simulation outputs the current social network (one file each time is does this!)
\end{itemize}

The following are scaling parameters.
\begin{itemize}
\item birth-mult: a scaling parameter that changes the birth rates uniformly
\item death-mult: a scaling parameter that changes the death rates uniformly
\item move-prob-mult: a scaling parameter that changes the probability of moving
\item influence-rate: a scaling parameter determining the maximum number of chances to influence others each agent has each year (this will be realised by very few agents if any, but will have the effect of scaling the number of discussions agents who are politically interested agents have)
\item forget-mult: a scaling parameter that changes the rate of forgetting
\end{itemize}


\end{document}